\documentclass[12pt]{article}

\usepackage{lineno}
\usepackage{xcolor}
\usepackage{svg}

\usepackage[utf8x]{inputenc} 
\usepackage{a4,color}
\usepackage{bm}
\usepackage{graphicx,psfrag}
\usepackage{dsfont}
\usepackage{amsmath,amssymb} 
\usepackage{caption}
\usepackage{subcaption}
\usepackage{hyperref}
\usepackage{comment}

\definecolor{purple}{rgb}{0.60, 0.00, 0.50}

\usepackage{algorithm}
\usepackage[noend]{algorithmic}

\usepackage{stmaryrd}
\usepackage{booktabs}
\usepackage{amsfonts}
\usepackage{fancyhdr}
\makeatletter
\newcommand*\dotp{\mathpalette\dotp@{.75}}
\newcommand*\dotp@[2]{\mathbin{\vcenter{\hbox{\scalebox{#2}{$\m@th#1\bullet$}}}}}
\makeatother

\usepackage{appendix}
\usepackage{xcolor}
\definecolor{dmblue500}{HTML}{0044CC}
\definecolor{dmorange500}{HTML}{FF5F19}
\definecolor{dmred500}{HTML}{E1144B}
\definecolor{dmpurple500}{HTML}{6932E6}
\definecolor{dmteal500}{HTML}{14C8B9}

\newcommand{\citep}{\cite}

\title{A Unified Theory of Dual-Process Control}

\author{\normalsize Ted Moskovitz\textsuperscript{1,2,*}, Kevin J. Miller\textsuperscript{2,3}, Maneesh Sahani\textsuperscript{1}, Matthew M. Botvinick\textsuperscript{1,2} 
\\
\\
{\normalsize 1. Gatsby Computational Neuroscience Unit, University College London, London, UK} \\
 {\normalsize 2. Google DeepMind, London, UK} \\
 {\normalsize 3. Department of Ophthalmology, University College London, London, UK} \\
{\normalsize *Correspondence: \texttt{ted@gatsby.ucl.ac.uk}}
}

\begin{document}
\maketitle

\abstract{Dual-process theories play a central role in both psychology and neuroscience, figuring prominently in domains ranging from executive control to reward-based learning to judgment and decision making. In each of these domains, two mechanisms appear to operate concurrently, one relatively high in computational complexity, the other relatively simple. Why is neural information processing organized in this way? We propose an answer to this question based on the  notion of compression. The key insight is that dual-process structure can enhance adaptive behavior by allowing an agent to minimize the description length of its own behavior. We apply a single model based on this observation to findings from research on executive control, reward-based learning, and judgment and decision making, showing that seemingly diverse dual-process phenomena can be understood as domain-specific consequences of a single underlying set of computational principles.
}

\subsection*{Introduction}

William James famously distinguished between two modes of action selection, one based on habit and the other involving effortful deliberation \citep{james1890principles}. This idea has since ramified into a variety of `dual-process' theories in at least three distinct domains of psychology and neuroscience. One of these domains concerns executive control, and  distinguishes action selection that is "automatic", reflecting robust stimulus-response associations, from that which is "controlled", overriding automatic actions when necessary \citep{diamond2013executive, botvinick2014computational}. A second focuses on reward-based learning, distinguishing behavior that is sensitive to current goals ("goal-directed" or "model-based") from that which is habitual \citep{dolan2013goals, perez2020theory}. The third addresses judgment and decision making (JDM), where canonical theories distinguish between two cognitive systems: a "System 1", which employs fast and frugal heuristic decision strategies, and a "System 2", which supports more comprehensive  reasoning \citep{evans2008dual, kahneman2011thinking}.  

Across these three domains, dual-process models  have accumulated considerable empirical support, and each domain has developed explicit computational models of how dual processes might operate and interact \citep{lieder2017strategy, botvinick2014computational, rougier2005prefrontal, daw2005uncertainty, shenhav2013expected, keramati2011speed, boureau2015deciding, perez2020theory, miller2019habits}. These computational models, however, are typically domain-specific, reproducing behavioral phenomena that are within the scope of their domain. It remains unknown whether dual-process phenomena in different domains result from different sets of computational mechanisms, or whether they can be understood as different manifestations of a single, shared set. That common mechanisms might be at play is suggested by a wealth of neuroscientific data. Specifically, studies have linked controlled behavior, model-based action selection, and System-2 decision making with common circuits centering on the prefrontal cortex \citep{diamond2013executive, dolan2013goals, mevel2019developmental, de2011heuristics, miller2001integrative, jeon2015degree} (Figures 2A, 3A, 5A, 6A), while automatic behavior, habitual action selection, and heuristic decision making appear to engage shared circuits lying more posterior and running through the dorsolateral striatum \citep{lieberman2007social, o2020sequential, jeon2015degree, smith2022habit}. 

In addition to reproducing dual-process phenomena across domains, a complete theory would provide a normative account explaining \textit{why} a dual-process computational architecture is adaptive. Why should human decision-making be organized in this form? Can dual-process decision-making be understood as a formally sound solution to a particular ethologically important computational problem \citep{musslick2021rationalizing, milli2021rational, maisto2019caching, ritz2022cognitive, piray2021linear}? 

In the present work, we address these open problems by offering a computational account of dual-process control that is both unifying and normative. We start by considering a fundamental challenge in adaptive behavior: the problem of generalization. Drawing on machine learning and information theory, we then show that a principled strategy for enhancing generalization leads directly to dual-process control. Finally, translating these insights into a runnable implementation, we demonstrate that a single computational model can explain canonical dual-process phenomena from executive control, reward-based learning, and JDM.  

\subsubsection*{Computational principle: Generalization via compression}

A fundamental demand of intelligent behavior is to capitalize on past learning in order to respond adaptively to new situations, that is, to generalize. Humans in particular show a remarkable capacity for behavioral generalization, to such a degree that this has been regarded as one of the hallmarks of human intelligence \citep{lake2017building}. Identifying the computational underpinnings of this ability stands as an important open problem. 

A useful context for thinking about generalization from a computational point of view is provided by the framework of reinforcement learning (RL; \citep{sutton2018reinforcement}). RL starts with an `agent' that receives observations of the environment and emits actions based on an adjustable `policy,' a mapping from situations to actions. For this agent, every situation is assumed to be associated with a quantitative reward. Based on its experience with actions and outcomes, the agent applies a learning algorithm to update its behavioral policy so as to progressively increase the amount of reward it collects \citep{sutton2018reinforcement}. Given these terms, generalization can be operationalized as an increase in the rate at which reward is received on a new task, attributable to previous exposure to one or more related tasks \citep{kirk2021survey}. Expressed equivalently, an agent generalizes effectively when it is able to quickly adapt so as to correctly predict, based on its past experience, which actions will be rewarding in some new situation.

Framing generalization in these predictive terms opens up a connection with the wider field of machine learning, where the problem of leveraging structure present in past data to predict future data constitutes a core disciplinary focus. In approaching this problem, the machine learning literature points consistently to the importance of \textit{compression}: In order to build a system that effectively predicts the future, the best approach is to ensure that that system accounts for past observations in the most compact or economical way possible \citep{mackay2003information, hutter2004universal, feldman2016simplicity, grunwald2007minimum}. One canonical method for specifying this compression objective more precisely is provided by the  \textit{minimum description length} (MDL) principle  \citep{grunwald2007minimum}. MDL theory proposes that the best representation or model $M$ for a body of data $D$ is the one that minimizes the expression
\begin{equation}
    L(M) + L(D|M).
\end{equation}    
$L(M)$ here is the description length of the model, that is, the number of information-theoretic bits it would require to encode that model, a measure of complexity \citep{grunwald2007minimum}. $L(D|M)$, meanwhile, is the description length of the data given the model, that is, an information measure indicating  how much the data deviates from what is predicted by the model. In short, MDL favors the model that best balances between deviation and complexity, encoding as much of the data as it can while also remaining as simple as possible.

\subsubsection*{Minimum description length control}

The MDL principle translates naturally into the context of RL. The key step is to designate, as the `data' to be compressed, the agent's behavioral policy (see Supplementary Discussion). Denoting this policy $\pi$, and following the logic of MDL, we also assume a `model' of the policy, which takes the form of an auxiliary policy $\pi_0$ (compare \cite{teh2017distral, tirumala2020behavior}). Following MDL further, we then define an optimization objective for both policies which weighs the standard RL term, favoring high expected reward ($R$), against the two terms of the MDL objective:
\begin{equation}
    E_{\pi}[R] - \lambda [ L(\pi_0) + L(\pi|\pi_0) ]
\end{equation}
with $\lambda$ as a weighting parameter. Maximizing this objective yields a form of regularized policy optimization which we will call \textit{minimum description length control}, MDL-C for short. The basic idea is to encourage the learning agent to formulate a policy that maximizes reward while also staying close to a simpler or more compressed reference policy. 

Recent advances in artificial intelligence (AI) allow us to implement MDL-C in the form of a runnable simulation model, as diagrammed in Figure 1 (see Methods). Here, both policy $\pi$ and policy $\pi_0$ are parameterized as identical recurrent neural networks, both receiving the same perceptual inputs. On every time-step, the network implementing the reference policy $\pi_0$ --- henceforth $RN\!N_{\pi_0}$ --- outputs a probability distribution over actions. That distribution is then updated by the network implementing policy $\pi$ ($RN\!N_\pi$), and the agent's overt action is selected (see Supplementary Discussion). 
To implement MDL regularization, the deviation term $L(\pi|\pi_0)$ is quantified as the Kullback-Leibler (KL) divergence between the two policies $\pi$ and $\pi_0$, consistent with the fact that the KL divergence represents the amount of information required to encode samples from one probability distribution (here $\pi$) given a second reference distribution ($\pi_0$). In order to implement the complexity cost $L(\pi_0)$, we apply a technique known as variational dropout (VDO; \citep{kingma2015vdo}). As detailed in the Methods section, VDO assumes that the synaptic weights in a neural network are subject to multiplicative Gaussian noise, and applies a form of regularization that biases toward high noise variance. Because as noise increases, the information carried by network weights decreases, VDO regularization can be understood as biasing networks toward compactness or simplicity (see \cite{hinton1993keeping, molchanov2017variational}). Combining both regularization terms with a standard RL reward objective results in a three-term objective function aligning with Eq. 2 (see Methods). Using this, the entire network is trained using a standard policy-gradient RL algorithm (see Methods).

Equipped with this runnable implementation, we can return to the problem of generalization, and ask whether MDL regularization in fact enhances generalization performance. In other words, we'd like to verify that this regularization enables the agent to adapt more quickly than it would otherwise to new goals. Figure 1B-C presents relevant simulation results (see also Methods, and \cite{moskovitz2023minimum} for related theoretical analysis and further empirical evaluation). When our MDL-C agent is trained on a set of tasks from a coherent domain (e.g., navigation or gait control) and then challenged with a new task from this same domain, it learns faster than an agent with the same architecture but lacking MDL regularization. In short, policy compression, following the logic of MDL, enhances generalization. 

Having established these points, we are now in position to advance the central thesis of the present work: We propose that MDL-C may offer a useful explanatory model for dual-process phenomena, as encountered in brain and behavior. As in dual-process theory, MDL-C contains two distinct decision-making mechanisms. One of these (corresponding to $RN\!N_{\pi_0}$ in Figure 1A) distills as much target behavior as possible in an algorithmically simple form, reminiscent of the habit system or System 1 in dual-process theory. Meanwhile, the other ($RN\!N_{\pi}$) enjoys greater computational capacity and intervenes when the simpler mechanism fails to select the correct action, reminiscent of executive control or System 2 in dual-process theory. MDL-C furnishes a normative explanation for this bipartite organization, by establishing a connection with the problem of behavioral generalization. 

This normative argument would gain additional force if MDL-C turned out also to provide a unifying perspective, identifying a common basis for observations spanning the three behavioral domains where dual-process theory has been principally applied. To test this, we conducted a series of simulation studies, each one applying our neural network implementation of MDL-C to a specific empirical domain: executive control in Simulation 1, reward-based decision making in Simulation 2, and JDM in Simulation 3.

\subsection*{General methods: Selection of target phenomena and approach to modeling}

A detailed description of simulation methods, sufficient to fully replicate our work, is presented in Online Methods. Briefly, for each target dual-process domain, we focused on a set of empirical phenomena that the relevant specialty literature treats as fundamental or canonical. We do not, of course, address all behavioral and neural phenomena that might be considered relevant to constrain theory in each domain, and we dedicate a later section to the question of whether any empirical findings that we do not directly model might present challenges for our theory. Nevertheless, the core phenomena in each field are fairly well recognized, and we expect our selections will be uncontroversial. Indeed, each target phenomenon has been the focus of previous computational work, and we dedicate a later section to comparisons between our modeling approach and previous proposals. While such comparisons are of course important, one point that we continue to stress throughout is that no previous model has addressed the entire set of target phenomena, bridging between the three domains we address.  

For each target phenomenon, we pursue the same approach to simulation: We begin with a generic MDL-C agent model, configured and initialized in the same way across simulations (with the exception of input and output unit labels tailored to the task context). The model is then trained on an appropriate target task and its behavior or internal computations queried for comparison with target phenomena. Importantly, the model is in no case directly optimized to capture target phenomena; the reported effects are always emergent, resulting from an interaction between the structure of the behavioral task and the logic of MDL-C. In the rare case where target effects depend sensitively on experimenter-chosen hyperparameters of MDL-C, this dependency is described alongside other results. 

While our simulations focus on target phenomena that have been documented across many experimental studies, in presenting each simulation we focus on observations from one specific (though representative) empirical study, to provide a concrete point of reference. It should be noted that the target phenomena we address, in almost all cases, take the form of qualitative rather than quantitative patterns. Our statistical tests, described in Online Methods, thus take the form of qualitative hypothesis tests rather than quantitative fits to data, paralleling the reference experimental research.

\subsection*{Results}

\subsubsection*{Simulation 1: Executive control}

\begin{figure}[t]
	\begin{center}
		\includegraphics[width=0.8\linewidth]{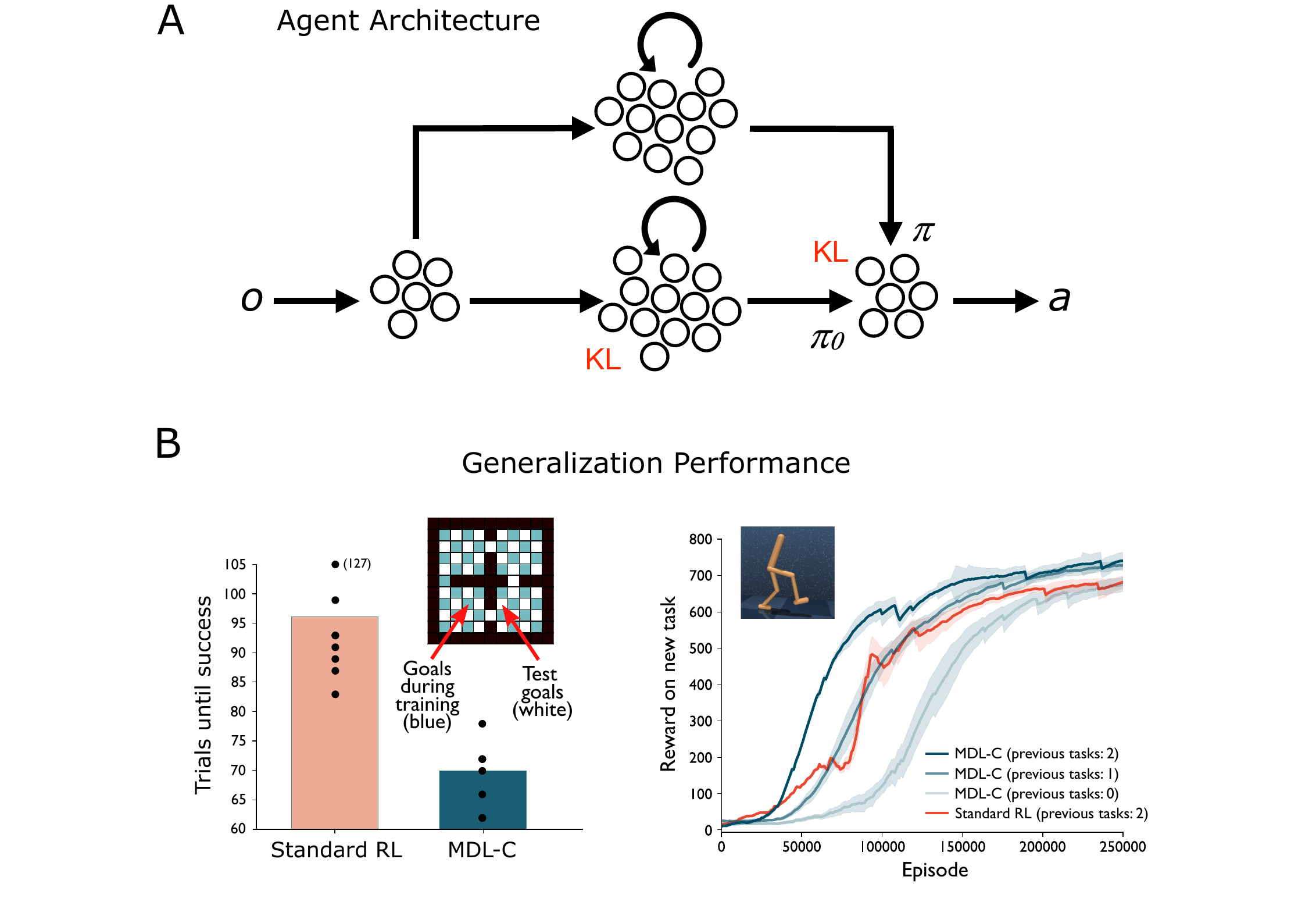}
	\end{center}
	\vspace{-2em}
	\caption{\scriptsize A. Schematic of a neural network implementation of MDL-C. Perceptual observations (input \textit{o} to units on the far left) feed into two recurrent networks. The lower pathway, $RN\!N_{\pi_0}$, contains noisy synaptic connections subject to VDO regularization as described in the main text. This network outputs an action distribution $\pi_0$. The upper pathway, $RNN_\pi$, outputs a separate action distribution $\pi$, which overwrites $\pi_0$, the KL divergence between the two policy outputs is computed, and an action \textit{a} is selected from $\pi$. Arrows indicate all-to-all projections. Not shown is a gating layer between input features and $RN\!N_{\pi_0}$, which was included for intepretability of results. See Methods for this and all other implementational details. B. MDL regularization enhances generalization. Left: Two agents, an MDL-C agent and an unregularized baseline (\textit{Standard RL}), were trained to navigate within a partitioned grid (inset) to a set of cued goal locations (blue tiles). In a second phase of training, the remaining (white) locations were presented as goals. The barplot shows the average number of trials elapsed before the agent first discovered a shortest path to goal. Individual points here and in subsequent figures are based on data from independent simulation runs. See Methods for details. Right: Average reward in a continuous control (running) task. MDL-C learns faster if the agent has previously encountered related tasks, and learns faster than a comparison agent lacking MDL-C's complexity penalty (\textit{Standard RL}; see Methods). Error bands here and in subsequent figures indicate standard error. In these and subsequent comparisons, unless otherwise noted, reported effects  were confirmed by rank-sum test at a threshold of p = 0.05 (see Methods).
	}
	\label{fig:fig1}
	\vspace{-1em}
\end{figure}

As introduced above, longstanding theories of executive function center on a contrast between two kinds of action. Habitual or automatic responses are default, reactive actions, shaped by frequency or practice.  Controlled responses, in contrast, take fuller account of the task context, overriding automatic responses when they are inappropriate \citep{diamond2013executive, botvinick2014computational, miller2001integrative}. Some of the strongest support for this distinction comes from studies of prefrontal cortex. Prefrontal neural activity has been shown to play a special role in encoding goals, task instructions, and other aspects of task context \citep{miller2001integrative, diamond2013executive}. The importance of these representations for context-appropriate behavior is evident in the effects of prefrontal damage, where behavior tends to default to frequently performed actions, neglecting verbal instructions or context-appropriate goals. 

One  domain in which these effects can be observed in a particularly straightforward form is spatial navigation (see Figure 2A). In navigation tasks, goal locations have been shown to be encoded in prefrontal cortex \citep{patai2021versatile}. And prefrontal damage impairs the ability to navigate to instructed goal locations, with behaviour defaulting to more familiar paths and destinations \citep{ciaramelli2008role} (Figure 2B).

\begin{figure}[t]
	\begin{center}
		\includegraphics[width=12cm]{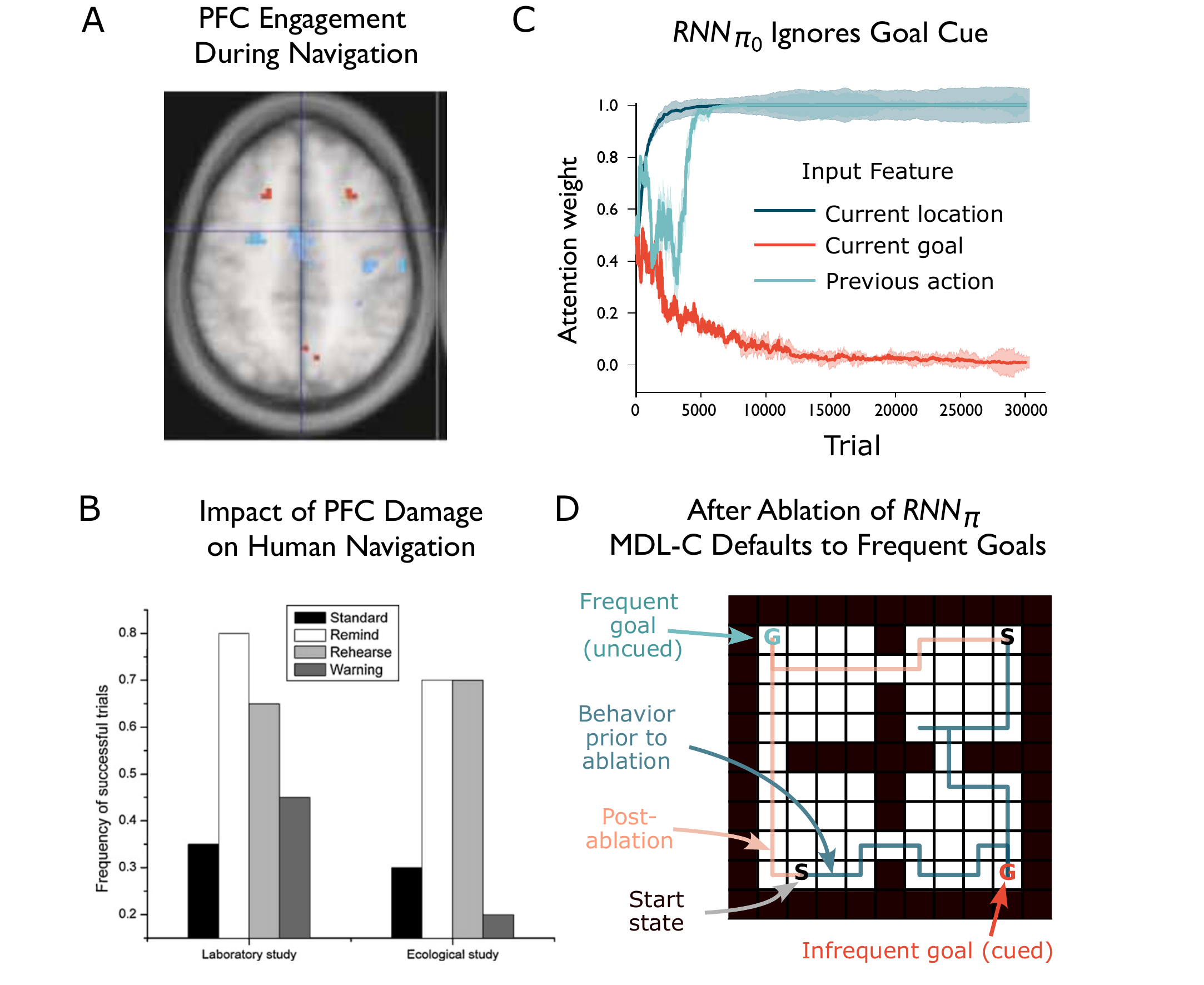}
	\end{center}
	\vspace{-2em}
	\caption{\scriptsize A. \cite{hartley2003well} reported greater activity in dorsolateral PFC (red) during wayfinding than during following of a familiar route, with the opposite effect in more posterior regions (blue). B. \cite{ciaramelli2008role} reported that damage to another (orbitofrontal) region of PFC impaired navigation to novel goals, both in the laboratory and an ecological study. In unsuccessful trials patients frequently navigated to familiar goal locations. Performance improved when patients were given frequent reminders of the goal or were asked to verbally rehearse the goal, but not when the goal reminder was replaced by an uniformative stimlus (\textit{Warning}). C. By inserting a gating layer over input features within $RN\!N_{\pi_0}$ (see Methods), we can directly read out which information is processed by that pathway. The plot shows attention weights for the three input features in the navigation task referenced in Figure 1. Over the course of the initial training block, $RN\!N_{\pi_0}$ learns to ignore the current goal cue. D. In a modified navigation task only two goals were cued, one (blue \textit{G}) occurring more frequently during training than the other (red \textit{G}). When the infrequent goal is cued at test, the intact MDL-C agent navigates successfully to it from any start state (see blue example trajectories). When $RNN_{\pi}$ is ablated, the agent ignores the instruction cue and navigates to the more frequent goal (pink trajectories). See Methods for simulation details. 
	}
	\label{fig:fig2}
	\vspace{-1em}
\end{figure}

Strikingly similar effects arise when MDL-C is applied to spatial navigation. In our first simulation, the MDL-C agent from Figure 1 was trained on a navigation task involving two cued goal locations, with one goal presented more frequently than the other (see Methods). Results are shown in Figure 2C-D. As in neuroscientific studies, following training, the current goal is represented in only one part of the agent, namely $RN\!N_{\pi}$. This network thus emergently assumes a functional role analogous to that of prefrontal cortex. And as seen following prefrontal damage, when $RN\!N_{\pi}$  is ablated, leaving behavior fully dependent on $RN\!N_{\pi_0}$, the agent ignores the goal cue, always heading toward the frequently visited default destination. $RN\!N_{\pi_0}$, in this sense, can be viewed as encoding habits, frequently performed action sequences that can be executed without guidance from explicit goal representations. All of these patterns arise despite the fact that $RN\!N_{\pi}$ and $RN\!N_{\pi_0}$ receive exactly the same external inputs, have the same number of units and connectivity, and are trained concurrently using a single optimization objective. 

To evaluate the generality of these effects, we applied MDL-C to  another classic executive control problem, the Stroop task \citep{stroop1935studies} (see Methods and Figure 3). Here, words that name colors are presented in hues that are either incongruent (e.g. \textit{RED} presented in blue) or congruent (\textit{RED} in red). An instruction cue indicates whether the current task is to read the word, the highly practiced automatic response, or to name the color, requiring cognitive control. According to neuroscientific models of Stroop performance, prefrontal cortex plays a special role in encoding task cues \citep{miller2001integrative, herd2006neural}, and this is consistent with the effects of prefrontal damage, which induces disproportionate impairments in responding to incongruent color-naming trials, where top-down control is most demanded \citep{tsuchida2013core} (Figure 3A). 

\begin{figure}[t]
	\begin{center}
		\includegraphics[width=12cm]{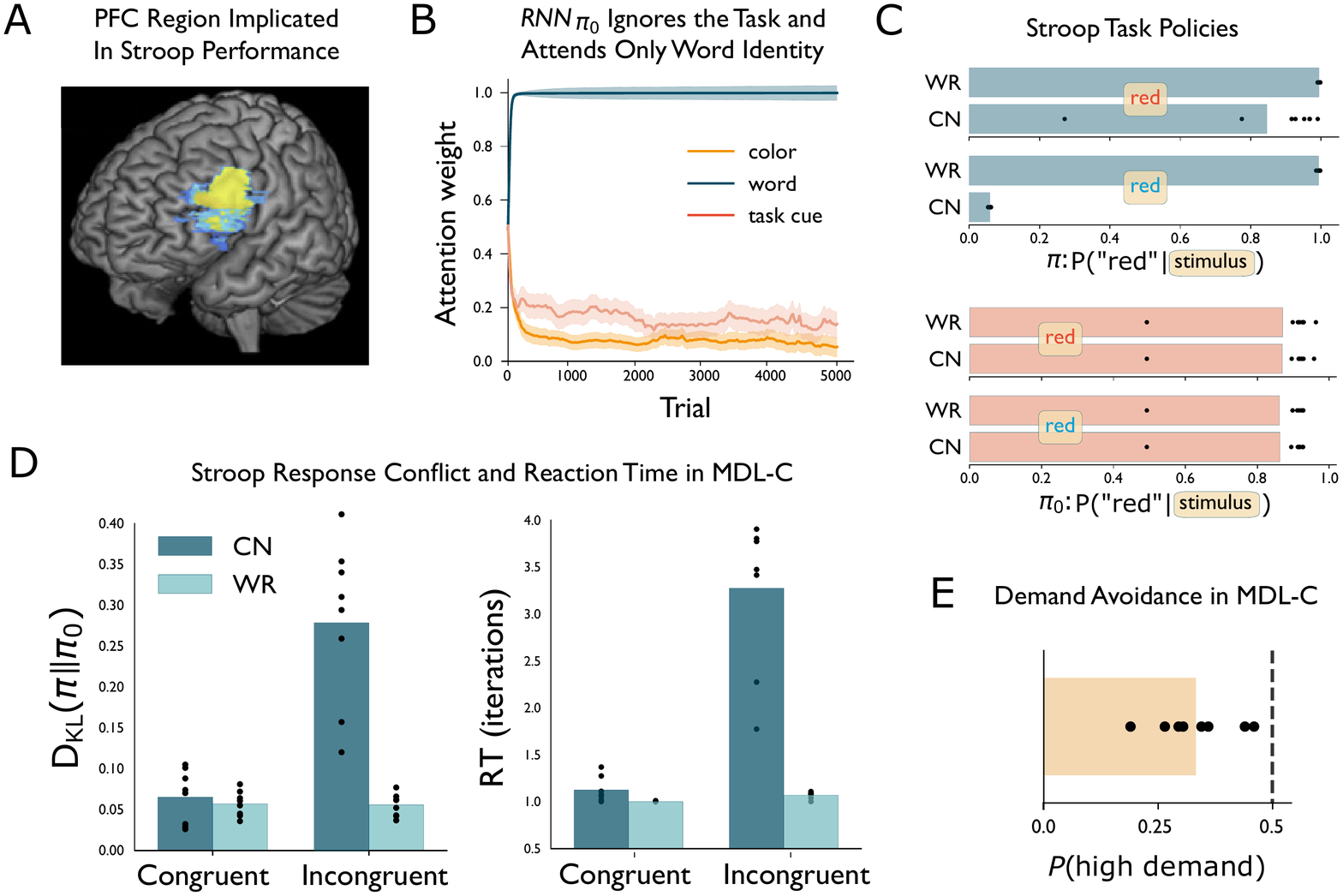}
	\end{center}
	\vspace{-2em}
	\caption{\scriptsize A. Consistent with many other findings, \cite{tsuchida2013core} reported an association between damage to dorsolateral PFC and Stroop interference. B. When the MDL-C agent is trained on the Stroop task (see Methods), $RN\!N_{\pi_0}$ learns to ignore both the task cue and the stimulus color, attending only to word identity. C. Policies for $RN\!N_{\pi}$ (top) and $RN\!N_{\pi_0}$ (bottom) for the stimuli shown, in word-reading (\textit{WR}) and color-naming (\textit{CN}) contexts. Response probabilities are shown for the response \textit{red}, complementary to (unshown) probabilities for the alternative \textit{blue} response.  D. Left: KL divergence between $\pi$ and $\pi_0$ for the four trial types shown in panel C. Right: Corresponding reaction times (see Methods). E. When trained on the Stroop task and then given a choice between blocks of color-naming trials that involve either high or low proportions of incongruent stimuli (see Methods), the MDL-C agent displays a preference for less frequent incongruence, paralleling the demand-avoidance effect seen in human decision making.   
	}
	\label{fig:fig3}
	\vspace{-1em}
\end{figure}

When trained on the Stroop task, MDL-C gives rise to precisely the same pattern of effects.  $RN\!N_{\pi}$, analogous to prefrontal cortex, represents the task instruction (Figure 3B). In contrast $RN\!N_{\pi_0}$, as in navigation, ignores the task context and is biased toward the behaviors executed most frequently during learning, consistent with the classical definition of automatic processing. This can be seen directly in the action distributions output by $RN\!N_{\pi_0}$, which look similar to those that would be appropriate in a word-reading context (see Figure 3C). In the intact agent, these habit-like responses are overridden (by policy $\pi$) only when the task context requires it. Such override events can be identified by tracking the KL divergence between policies $\pi$ and $\pi_0$. As shown in Figure 3D, this KL is highest precisely on incongruent color-naming trials. The overall pattern of KL divergences closely resembles the agent's reaction times, which in turn reproduce the patterns seen in human behavioral studies of the Stroop task and addressed in many previous computational models (see \cite{botvinick2014computational, herd2006neural}). This is especially interesting as one of the oft-cited normative advantages of habits is that they are ``faster.'' Here, we see that, indeed, lower reaction times are associated with selecting habitual actions.

In summary, MDL-C reproduces the canonical findings from empirical studies of automatic and controlled processing, across disparate tasks. Training with the MDL-C objective leads emergently to a habit system, which implements the most frequently performed behaviors, and a control system that captures additional aspects of context and overrides habits when this is demanded by context.  

Importantly, in addition to matching these points, MDL-C also offers a normative computational explanation for the fact that  brain function is organized in this way. Specifically, the dual-process pattern of functional differentiation arises through learning in MDL-C because it minimizes the description length of the agent's policy. This is shown more directly in Figure 4, which indicates how the total description length of the policy changes as we re-scale the complexity penalty imposed on $RN\!N_{\pi_0}$ during training (see Methods). At the right of the displayed plots, the description length is relatively high because $\pi_0$ is essentially uniform, requiring $\pi$ to depart frequently from $\pi_0$, incurring large KL costs. At the left of the plot, $\pi_0$ almost always selects the correct action, but only by building in excessive algorithmic complexity. The sweet spot lies near the middle of the plot, where $\pi_0$ absorbs the most frequent behaviors while remaining simple, and $\pi$ needs to override $\pi_0$ only infrequently. Our proposal is that executive control and habit systems in the brain strike the same balance, effectively minimizing the description length of adaptive behavior. 

\begin{figure*}[t]
	\begin{center}
		\includegraphics[width=10cm]{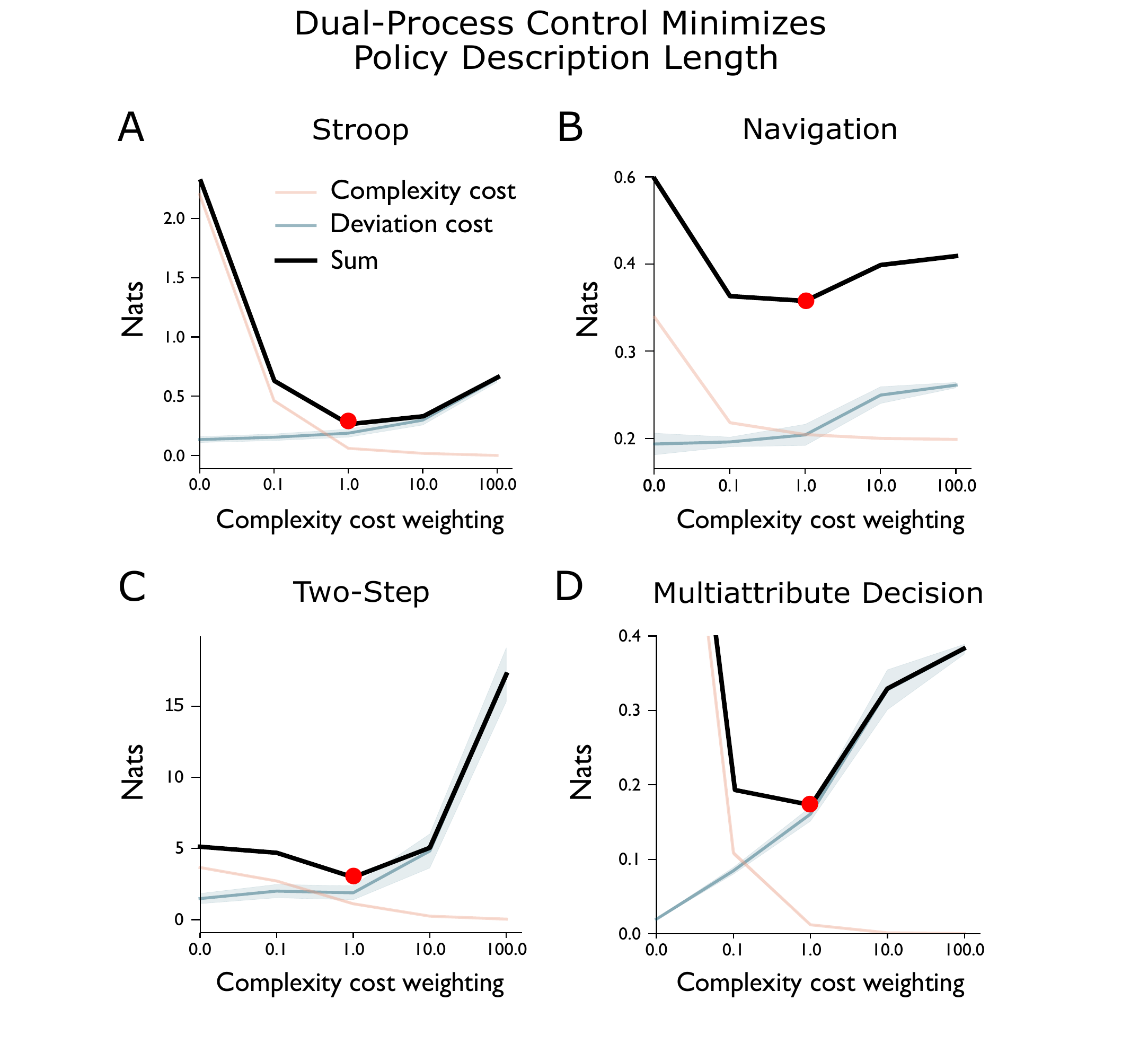}
	\end{center}
	\vspace{-2em}
	\caption{\scriptsize MDL-C minimizes the description length of its own behavior. In each plot, the x axis tracks the multiplicative weight placed on the complexity term in the MDL objective during training (see Methods). The y axis shows the KL costs, expressed in natural units of information (nats), corresponding to the deviation term in the MDL-C objective (blue), the complexity term (pink) and their unweighted sum (black). The complexity term naturally falls with increases to the weight placed on the complexity penalty during training. At the same time, this reduction in complexity causes the policy $\pi_0$ to diverge from the policy $\pi$, progressively inflating $D_{KL}(\pi || \pi_0)$. Note that the KL sum is the quantification of the full description length that applies when the complexity cost weight is 1.0 (see Methods and Supplementary Discussion). As shown, this measure displays a U-shaped profile with a minimum at 1.0. MDL-C thus minimizes the description length of behavior, as quantified within the objective function employed for training. Panels A-D illustrate the effect for our simulations of the Stroop task (A); navigation as referenced in Figure 1 (B); the two-step task (C); and the multiattribute decision-making task from Simulation 3, in the parameter regime producing behavior as in Figure 6D (D).
	}
	\label{fig:fig4}
	\vspace{-1em}
\end{figure*}

As it turns out, the description-length principle also accounts for one other core phenomenon in the cognitive control literature, namely \textit{demand avoidance}, the tendency for decision makers to avoid tasks that require intensive cognitive control \citep{kool2018mental}. For example, when human participants are asked to select between two versions of the Stroop task, one involving more frequent incongruent trials than the other, they show a clear tendency to avoid the former task and the demands on cognitive control it involves  \citep{schouppe2014context}. When MDL-C is trained in the same task context (see Methods), the same choice bias arises (Figure 3E). The explanation for this result is tied to the final term in the MDL-C objective function (see Equation 2), which penalizes conflict between policies $\pi$ and $\pi_0$ (compare \cite{zenon2019information, piray2021linear}). By avoiding control-demanding tasks, the agent can minimize this term, helping it to minimize the description length of its overall behavioral policy. 

The relation of the above simulation results to those from previous models, and a consideration of a wider range of empirical phenomena in the domain of executive control, are discussed below under \textit{Comparison with previous models}.   

\subsubsection*{Simulation 2: Reward-based learning}

\begin{figure*}[t]
	\begin{center}
		\includegraphics[width=12.0cm]{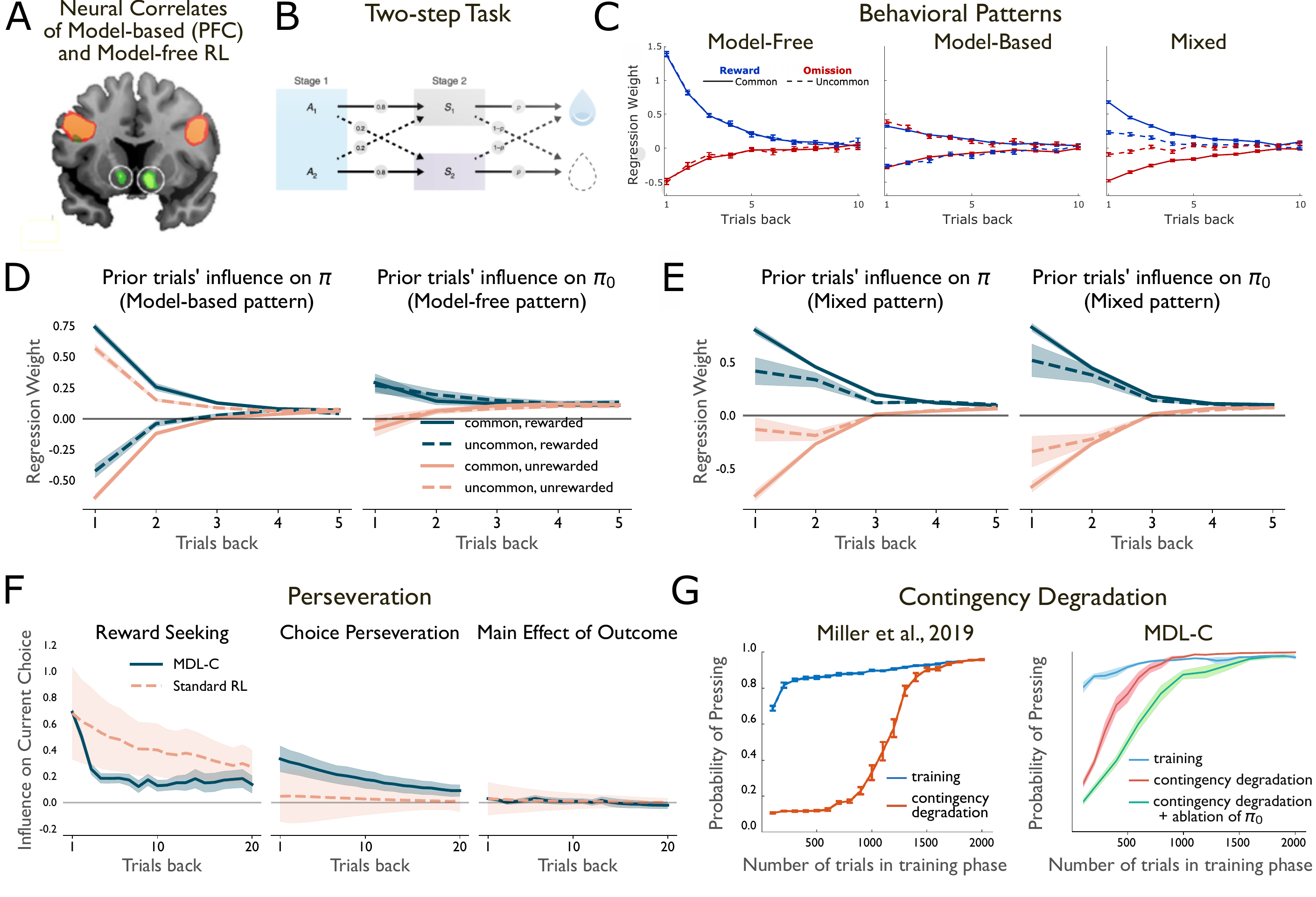}
	\end{center}
	\vspace{-2em}
	\caption{\scriptsize \textbf{A}. In line with many other pieces of evidence, \cite{glascher2010states} reported a functional-anatomical dissociation between signals related to model-based RL, in dorsolateral PFC (orange), and signals related to model-free RL (green). \textbf{B}. Structure of the two-step task as introduced by \cite{daw2011model}. Choice occurs at Stage 1. The value of \textit{p} varies over time, and so must be inferred by the participant. Following subsequent research, the version employed in our experiments additionally included explicitly cued reversals in the structure of transitions from Stage 1 to Stage 2. See Methods for full details. \textbf{C}. Classical behavioral signatures of model-free (left) and model-based (center) performance in the two-step task. Adapted from \cite{miller2016identifying}, the plots show logistic regression weights quantifying the influence of two factors on the probability of repeating on the index trial the same first-stage action selected on the previous trial: (1) whether reward was received or omitted on the previous trial, and (2) whether the previous trial featured a transition from stage 1 to 2 that was high-probability (\textit{common}) or low (\textit{uncommon}). The right panel shows a hybrid pattern, similar to that reported in the classic study by \citep{daw2011model}. \textbf{D}. Left: Two-step behavior of MDL-C, reflecting policy $\pi$. Right: Influence of the past on policy $\pi_0$. 
	\textbf{E}. Same as Panel D but with different weighting of terms in the MDL-C objective (see Methods and compare panel C, right). \textbf{F}. Logistic regression weights showing the influence on the current action of reward contingent on choice (reward seeking), previous choices (perseveration), and reward independent of choice (main effect of outcome) of MDL-C and a standard RL agent on the drifting two-armed bandit task from \cite{miller2018predictive}. MDL-C displays a stronger tendency towards perseveration, reminiscent of rats trained on the same task. \textbf{G}. Left: Simulation of contingency degradation from \cite{miller2019habits}. The longer the training phase (x axis), the longer lever-pressing persists after reward is discontinued (red). Right: Corresponding behavior from MDL-C, also showing the effect of ablating $\pi_0$.
	}
	\label{fig:fig5}
	\vspace{-1em}
\end{figure*}

According to prevailing theories, reward-based learning centers on two distinct neural systems (Figure 5A). One, operating within parts of prefrontal cortex and associated basal ganglia circuits, implements a `goal-directed' or `model-based' algorithm, which takes task structure into account. The other system, more posterior or lateral, operates in a `habitual' manner, based on simpler stimulus-response associations \citep{dolan2013goals, daw2005uncertainty, beierholm2011separate, glascher2010states, averbeck2022reinforcement, drummond2020model, miller2019habits, dickinson1985actions}. 
Although the anatomical substrates proposed for these systems can resemble those associated with controlled and automatic processing,  different behaviors have been used to study them. In research with humans, the most prominent of these is the so-called `two-step task' \citep{daw2011model}, illustrated in Figure 5B-C. 

The two-step task was designed to probe the operation of model-based and habitual systems, under the hypothesis that these operate in parallel and that the habitual system implements model-free reinforcement learning \citep{averbeck2022reinforcement, drummond2020model} (see \textit{Comparison with previous models} and Supplementary Discussion). We focus on a variant of the task designed to maximize the task's ability to uncover such computational structure (Figure 5B and Methods). When we train a MDL-C agent under certain parameterizations, we find the division of labor described in the dual-process literature (Figure 5c, left and center): the patterns associated with model-based and model-free control arise side by side, with policy $\pi$ displaying the model-based profile, and $\pi_0$  the model-free pattern (Figure 5D). Because $\pi$ dictates the overt behavior of the agent, the latter displays a model-based pattern, as also seen in human performance in some studies \citep{feher2020humans}. When $RN\!N_{\pi}$ is ablated, behavior then shifts away from the model-based pattern, in line with the observation that disruption of prefrontal function decreases model-based control in the two-step task \citep{smittenaar2013disruption, otto2013curse}.

This differentiation of function arises, as in the previous simulations, from the MDL-C optimization objective. As has been noted in the literature on model-based versus model-free learning, the latter is less algorithmically complex \citep{daw2005uncertainty}. The simplicity bias in MDL-C, imposed on $\pi_0$, therefore tilts that policy toward the actions that would be chosen by a model-free agent. Policy $\pi$, meanwhile, can reap a bit more reward by implementing a policy that takes task structure more fully into account. The overall division of labor minimizes the description length of behavior, in line with the illustration in Figure 4C.

While MDL-C thus captures the sharp functional contrast proposed by theory, it can also address empirical data suggesting a more nuanced dual-system division of labor. Specifically, experimental findings indicate that human subjects typically display behavior that appears intermediate between the classical model-based and model-free patterns (see Figure 5C, right, and \cite{daw2011model}). In MDL-C, when the parameters weighting the terms in the objective function are varied (see Methods), such a hybrid pattern is observed across large portions of the parameter space (Figure 5E and Supplementary Discussion). Thus, while a clean separation between model-based and model-free learning can arise within MDL-C, such a division is not hardwired into the framework. Depending on the precise setting, minimizing the description length of behavior can also lead to graded intermediate patterns, providing leverage on some otherwise problematic experimental observations \citep{collins2020beyond}.

While the two-step task has been an important driver of dual-process theory in the domain of reward-based learning, important insights have also come from studies of instrumental learning. Within this literature, one particularly important experimental manipulation is known as \textit{contingency degradation}. Here, rewards are at first delivered only in response to a particular action, but then later are delivered in a non-contingent manner, independent of whether the action was selected. Unsurprisingly, this change typically triggers a shift away from the action in question. Critically, however, this adjustment is reduced or slowed if the initial training with reward was extensive \citep{daw2005uncertainty, miller2019habits, dickinson1998omission} (Figure 5G). Prevailing explanations for this effect share a dual-process perspective, according to which insensitivity to contingency degradation reflects a transfer of control from one learning process that is relatively flexible to another which adjusts less quickly \citep{daw2005uncertainty, miller2019habits}. Consistent with this account, lesions to dorsolateral striatum --- a structure proposed to be involved in that latter system --- partially protects against training-induced inflexibility \citep{yin2006inactivation}.

MDL-C captures the empirically observed effects of contingency degradation, but also offers a novel computational perspective. As shown in Figure 5G, the speed with which the MDL-C agent reduces its response rate after contingency degradation depends on how long the agent was previously trained with reward (see Methods for simulation details). As in the experimental data, behavior becomes less flexible as the duration of training increases. This shift is an emergent result of the MDL-C optimization objective. Policy $\pi$ is initially able to adjust rapidly, responding to reward by emitting the rewarded action frequently. If contingency degradation occurs immediately, $\pi$ is able to adapt flexibly. However, if reward continues for a longer period, the rewarded policy gradually comes to be mirrored in $\pi_0$, driven by the third term in Equation 2. Once $\pi_0$ becomes strongly biased toward the rewarded action, it is difficult for policy $\pi$ to diverge from this pattern, again due to the third term in Equation 2 (an effect that is attenuated if $\pi_0$ is ablated, analogous to lesioning dorsolateral striatum; see Figure 5G). This computational mechanism is related to others that have been proposed in models devised specifically to account for contingency degradation effects, based on uncertainty or habit strength \citep{daw2005uncertainty, miller2019habits} (see Supplementary Discussion). However, MDL-C ties the relevant learning dynamics to a higher-level computational objective, namely, minimizing the description length of behavior (compare \cite{pezzulo2018hierarchical, lai2021policy}).

\subsubsection*{Simulation 3: Judgment and decision making}

\begin{figure*}[t]
	\begin{center}
		\includegraphics[width=12cm]{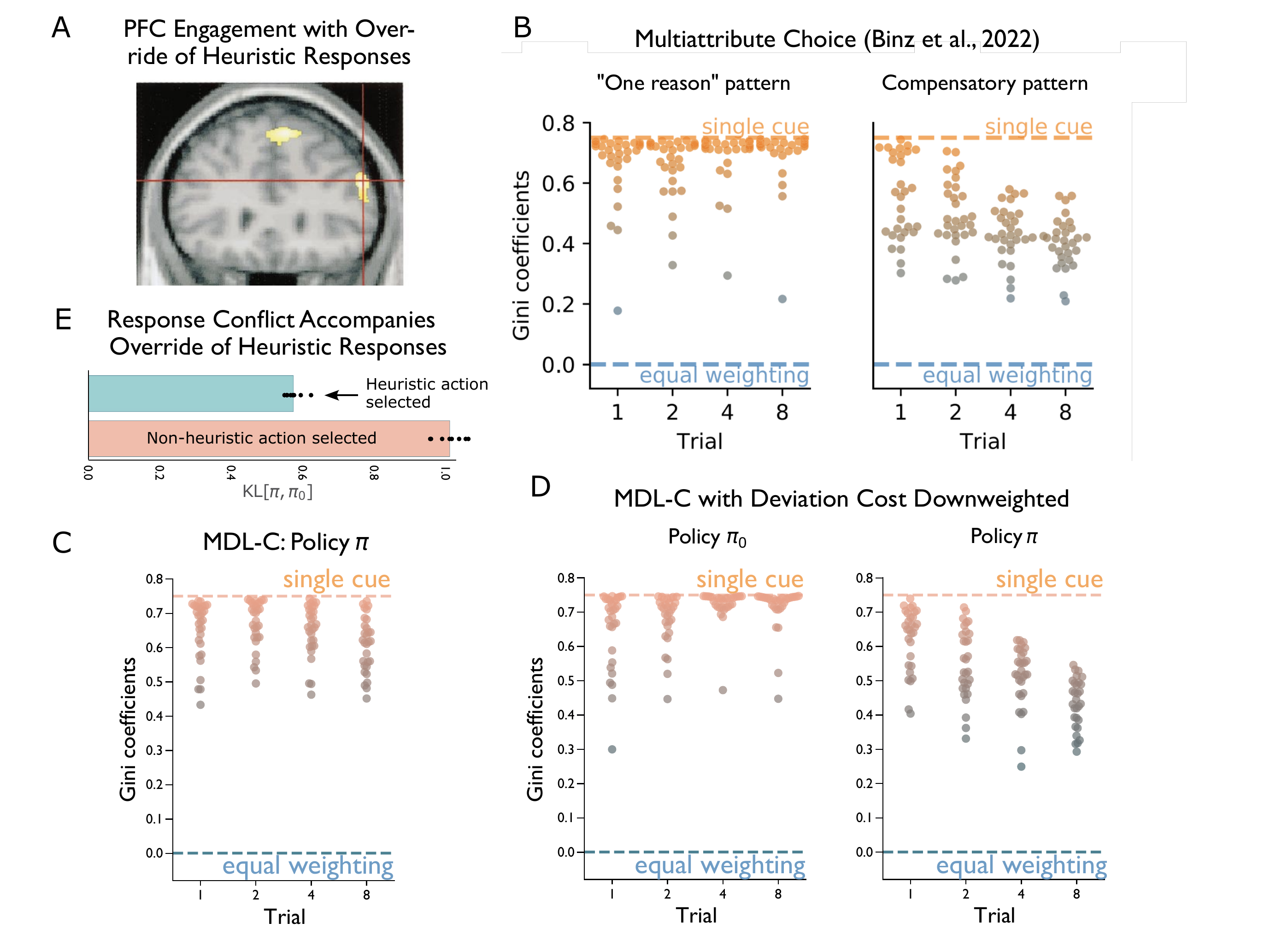}
	\end{center}
	\vspace{-2em}
	\caption{\scriptsize A. \cite{de2011heuristics} reported elevated activity in dorsolateral PFC when participants overrode heuristic-based responses to engage in richer reasoning. Image displays region of interest, drawn from \cite{goel2000dissociation}. B. Heuristic one-reason decision making (left) and richer compensatory decision making (right) in a multi-attribute choice task, from \cite{binz2022heuristics}. Gini coefficients, on the y axis, capture the degree to which decisions depend on one feature (higher values, with asymptotic maximum of one) versus all features evenly (zero), with references for one-reason decision making (\textit{single cue}) and a fully compensatory strategy (\textit{equal weighting}) indicated. Data points for each trial correspond to observations from separate simulation runs. Human participants in the study displayed both patterns of behavior, depending on the task conditions. C. Behavior of MDL-C in the task from \cite{binz2022heuristics}, under conditions where human participants displayed one-reason decision making. D. Behavior of $\pi_0$ (left) and $\pi$ (right) when the KL penalty for divergence between the two policies is reduced (see Methods). E. In the simulation from panel D, the divergence between policies is increased when the agent emits a non-heuristic decision. 
	}
	\label{fig:fig6}
	\vspace{-1em}
\end{figure*}

As noted earlier, dual-process models in JDM research distinguish between System-1 and System-2 strategies, the former implementing imprecise heuristic procedures, and the latter sounder but more computationally expensive analysis \citep{evans2008dual,kahneman2011thinking}. As in the other dual-process domains we have considered, there appears to be a neuroanatomical dissociation in this case as well, with System-2 responses depending on prefrontal computations \citep{mevel2019developmental, de2011heuristics}. 

Recent research on heuristics has increasingly focused on the hypothesis that they represent resource-rational approximations to rational choice \citep{lieder2020resource}. In one especially relevant study, \cite{binz2022heuristics} proposed that heuristic decision making arises from a process that ``controls for how many bits are required to implement the emerging decision-making algorithm" (p. 8).  This obviously comes close to the motivations behind MDL-C. Indeed, \cite{binz2022heuristics} implement their theory in the form of a recurrent neural network, employing the same regularization that we apply to our $RN\!N_{\pi_0}$. They then proceed to show how the resulting model can account for heuristic use across several decision-making contexts. One heuristic they focus on, called \textit{one-reason decision making}, involves focusing on a single choice attribute to the exclusion of others \citep{newell2003take}. As shown in Figure 6B, reproduced from \cite{binz2022heuristics}, a description-length regularized network, trained under conditions where one-reason decision making is adaptive (see \cite{binz2022heuristics} and Methods), shows use of this heuristic in its behavior, as also seen in human participants performing the same task. In contrast, an unregularized version of the same network implements a more accurate but also more expensive `compensatory' strategy, weighing choice features more evenly. 

As illustrated in Figure 6C, when MDL-C is trained on the same task as the one used by \cite{binz2022heuristics} (see Methods), it displays precisely the same heuristic behavior those authors observed in their human experimental participants. 

Digging deeper, MDL-C provides an explanation for some additional empirical phenomena that are not addressed by \cite{binz2022heuristics} or any other previous computational model.  In an experimental study of one-reason decision making, \cite{newell2003take} observed that application of the heuristic varied depending on the available payoffs. Specifically, heuristic use declined with the relative cost of applying a compensatory strategy, taking more feature values into account. MDL-C shows the same effect. When the weighting of the deviation term $D_{KL}(\pi || \pi_0)$ is reduced relative to the value-maximization term in the MDL-C objective (see Methods), the policy $\pi$ and thus the agent's behavior take on a non-heuristic compensatory form (Figure 6D). Critically, in this case MDL-C instantiates the non-heuristic policy side-by-side with the heuristic policy, which continues to appear at the level of $\pi_0$. This aligns with work suggesting that System-1 decision making can occur covertly even in cases where overt responding reflects a System-2 strategy. In particular, \cite{mevel2019developmental} observed activation in prefrontal areas associated with conflict detection in circumstances where a tempting heuristic response was successfully overridden by fuller reasoning (see also \cite{de2011heuristics}). A parallel effect is seen in our MDL-C agent in the degree of conflict (KL divergence) between policies $\pi$ and $\pi_0$ (Figure 6E).

\subsubsection*{Comparison with Previous Models}

\begin{figure*}[t]
	\begin{center}
		\includegraphics[width=10cm]{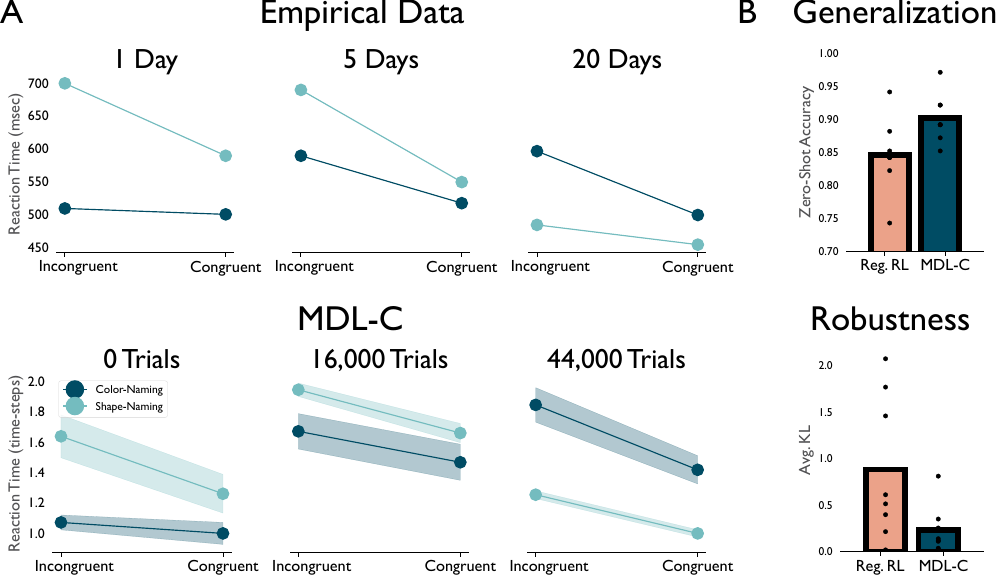}
	\end{center}
	\vspace{-2em}
	\caption{\scriptsize \textbf{A.} Top: Behavioral data from the modified Stroop task studied by \cite{cm1988training}. Early in training, shape-naming responses were  both slower than color-naming responses and more affected by stimulus congruence, consistent with shape-naming being the relatively 'controlled' response and color-naming relatively 'automatic.' With extensive training, the pattern flipped, with shape-naming becoming faster than color-naming and less affected by stimulus congruence.  Bottom: Under training conditions mimicking the experimental study, MDL-C displayed a similar pattern of behavior, with a significant main effect of task and a significant interaction between task and trial-type (p < 0.05) at both 0 trials and 44,000 trials. \textbf{B.} Zero-shot Stroop performance in MDL-C and an unregularized baseline model (see Methods). Top: Color-naming accuracy on incongruent Stroop stimuli, after training only with neutral stimuli (see main text and Methods). Bottom: KL divergence between action probability distributions under two conditions, (1) presentation of incongruent Stroop stimuli, and (2) presentation of Stroop stimuli with the word identity input masked out. MDL-C shows significantly lower divergence, indicating that the control policy attends less to the task-irrelevant factor --- i.e., MDL-C is more robust to distractors --- despite never having been trained on incongruent stimuli.
	}
	\label{fig:fig7}
	\vspace{-1em}
\end{figure*}

To our knowledge, no previous computational model has simultaneously captured the core dual-process phenomena we've considered, thereby bridging the domains of executive function, reward-based decision making and JDM. However, a range of previous models have addressed the relevant phenomena in a fashion limited to one of those domains. Having stressed the unifying, cross-disciplinary character of the present work, it is also befitting to consider the relationships between MDL-C and these domain-specific models. Particularly important is the question of whether such domain-specific models capture any empirical phenomena that MDL-C might have difficulty addressing.

In the area of executive control, our model bears strong connections with the classic connectionist model proposed by \cite{miller2001integrative}. In particular, both characterize the distinction between controlled and automatic processing as arising from learning.  To illustrate this point, \cite{cohen1990stroop} modeled results from a behavioral study by \cite{cm1988training} (Figure 7a). Here, participants were presented with colored shapes, and asked either to name their color or to announce a color name that had been arbitrarily assigned to the relevant shape (e.g., a particular irregular pentagon might be given the name \textit{blue}, independent of its display color). Interference between the two tasks was quantified by comparing response time on incongruent trials, where color- and shape-name conflicted, against congruent trials, where they matched.  Early in training, interference was larger for the shape-naming task than the color-naming task, suggesting that color-naming was relatively ‘automatic’ and shape-naming relatively ‘controlled.’  However, after extensive training on the shape-naming task the pattern flipped, consistent with the idea that within-task learning had rendered shape-naming relatively ‘automatic.’  This effect was well captured by the neural network model of \cite{cohen1990stroop}, and it also arises in our MDL-C model (see Figure 7a and Methods). 

As this example illustrates, gradual learning processes, operating over the course of extensive practice on relevant tasks, are important to the theoretical account we are proposing with MDL-C. On the face of it, this may seem to stand in tension with how learning occurs in most human behavioral experiments, where participants dive in on novel tasks given little more than some verbal instructions and few practice trials. For example, in the classic Stroop task, it seems reasonable to assume that participants have rarely if ever been asked to name the color of a word that itself names a color, but they do this ‘zero-shot,’ and immediately display Stroop interference. To show that our MDL-C implementation accommodates this kind of zero-shot learning, we trained our agent on color-naming and on word-reading, only ever presenting ‘neutral’ stimuli, omitting the word feature during color-naming and omitting the color feature during word-reading (see Online Methods). At test, incongruent feature sets were presented. The model responded correctly on the vast majority of trials given the task-cue input --- performing significantly better than an ablated network lacking MDL regularization --- but also showed Stroop interference (see Figure 7b).  In recent work, \cite{riveland2022neural} have shown how neural networks can follow verbal instructions zero-shot in a wider range of tasks. It would be exciting to expand our MDL-C implementation to incorporate greater behavioral breadth and flexibility in the same way.

Elaborations of the \cite{miller2001integrative} model have offered a mechanistic explanation for the special role played by prefrontal cortex in representing aspects of context, attributing to prefrontal circuits a special set of gating-based memory mechanisms \citep{o2010computational}. MDL-C offers a complementary account, instead addressing why it makes sense in normative terms for the brain to support both control and habit systems (see  \cite{musslick2021rationalizing} for a related but domain-specific analysis). As it turns out, however, MDL-C does in fact give rise to a solution that gates different information into different parts of the information-processing architecture, broadly consistent with gating-based models of cognitive control \citep{o2010computational}. From the point of view of our theory, such gating mechanisms might be viewed as solutions to the MDL-C problem discovered by evolution rather than online learning. It is worth noting that some of the most recent work to apply the notion of gating to PFC function has postulated a multilevel hierarchy, deeper than the one we consider in our simulations. There is no practical impediment to extending the MDL-C architecture to include multiple hierarchical levels; a natural approach would be to regularize each pair of adjacent layers with respect to one another, varying the weight of the complexity cost monotonically across layers.  We have not, however, implemented this idea and it therefore stands as an appealing opportunity for next-step research. 
Another elaboration of the \cite{miller2001integrative} model adds a `cost of control,' a negative utility attached to the overriding of default response-selection processes \citep{shenhav2013expected, zenon2019information, lieder2018rational, piray2021linear}. As noted in our simulation of demand avoidance, the deviation term in the MDL-C objective effectively imposes a cost of control, showing how this cost fits into a broader optimization process.

The classic \cite{miller2001integrative} model has been elaborated in subsequent work to address another canonical phenomenon in the executive function literature, which we have not previously touched upon: task-switching costs (see, e.g., \cite{herd2014neural}, \cite{reynolds2006computational}, \cite{gilbert2002task}). Importantly, in order to capture switch-cost effects, including such phenomena as residual and asymmetric switch costs, the relevant computational models have had to build in temporally and mechanistically fine-grained accounts of working memory function, modeling attractor dynamics and hysteresis effects that fall well below the level of abstraction our MDL-C implementation occupies. It would be informative to implement MDL-C with an increased level of temporal granularity (as for example in \cite{herd2014neural}) and to evaluate task-switching effects in this setting. 

We turn now from executive function to reward-based decision making. As shown in Simulation 2, when MDL-C operates within an appropriate task context, it can yield side-by-side decision mechanisms with profiles matching model-based and model-free control. This links MDL-C with a wide range of recent models of reward-based decision making, which center on this side-by-side configuration \citep{dolan2013goals, daw2005uncertainty, glascher2010states, beierholm2011separate}. As discussed under Results, the empirical data motivating those dual-system models is complex. In particular, neural activity aligning with model-free computations is not always `pure' of model-based characteristics (see, e.g., \cite{daw2011model}). Such computational purity is not enforced in MDL-C, either, and under some parameterizations MDL-C displays the same intermediate patterns that have been observed in some experimental studies. (Indeed, such mixed patterns were seen across most of the parameter space we explored; see Supplementary Figures 1-3). The interpretation of ostensibly model-based behavior in the two-step task is also nuanced \citep{akam2015twostep, miller2016identifying}. However, we have demonstrated elsewhere \citep{wang2018metapfc} that genuinely model-based computations can arise within recurrent neural networks under conditions comparable to those employed in the present work. 

Beyond model-based and model-free RL, the dynamics of habit acquisition in MDL-C also link it with recent models that replace model-free RL with a reward-independent, practice-based learning mechanism \citep{ashby2007neurobiological, miller2019habits,  bogacz2020dopamine}.  The learning mechanism of MDL-C's default policy is closely related to these, with two important differences. The first is that the practice-based learning mechanisms adopt as the target of learning the discrete actions actually taken by the agent, while MDL-C's default policy adopts as its target the full probabilistic control policy from which those actions are sampled. The second is that the addition of VDO effectively regulates the complexity of the habits that can be learned and the rate at which habit formation occurs. The results presented in Figure 5G support this connection. Of particular interest, a recent study provided evidence that dopamine dynamics in a posterior sector of the striatum encode not a reward-prediction error, but instead an \textit{action}-prediction error, which drives situation-action associations \citep{Greenstreet2022.09.12.507572}. This aligns quite closely with how learning operates in $RN\!N_{\pi_0}$ in our MDL-C implementation, where weight updates are driven by a mismatch between the actions predicted by $\pi_0$ and those dictated by $\pi$. 

Practice-based accounts of habits have been proposed \citep{miller2019habits} to explain not only classic assays of habits, but also trial-by-trial perseveration, an effect in which subjects tend to repeat in the future choices that have been made in the past, regardless of the associated stimuli and outcomes \citep{cho2002mechanisms, akaishi2014autonomous, balcarras2016attentional, miller2018predictive}. To test whether MDL-C would show such effects, we ran it on a drifting two-armed bandit task, in which rats show robust perseveration \citep{miller2018predictive}. We find that MDL-C shows similar perseveration, while an ablation model lacking the default policy does not (Figure \textbf{5F}). 

Despite all of these connections, MDL-C differs from most previous models in that it does not involve a direct competition between control systems \citep{daw2005uncertainty, lee2014neural}. In MDL-C, the policy $\pi$ always has the last word on action selection, which may be to either endorse or override default policy $\pi_0$ (as discussed above). Interestingly, this relationship between systems resembles one proposal for the interplay between System 1 and System 2 in the JDM literature, according to which ``System 1 quickly proposes intuitive answers to judgment problems as they arise, and System 2 monitors the quality of these proposals, which it may endorse, correct or override" \citep{kahneman2002representativeness}.

Within the JDM literature, among computational models of heuristic judgment, our account aligns closely with the one recently proposed by \cite{binz2022heuristics}, adding to it in the ways noted earlier. Like \cite{binz2022heuristics}, we have only applied MDL-C to a small set of heuristics from among the many considered in the JDM literature. An important challenge, both for MDL-C and for the \cite{binz2022heuristics} account, will be to test applicability to a wider range of the relevant behavioral phenomena. Needless to say, a still wider range of decision effects addressed by the JDM literature, from risk attitudes to self-control conflicts, remain untouched by the present introductory work, and the compatibility of the our theory with such effects will necessarily await further research. 

Some readers will have remarked that the our account of dual-process control  shares important characteristics with a range of research on `resource-rational' cognition \citep{lieder2020resource}, where limitations on computational capacity are understood to constrain strategies for adaptive information processing. In the context of goal pursuit, this perspective has given rise to the notion of a value-complexity tradeoff, where reward maximization balances against the cost of encoding or computing behavioral policies \citep{amir2020value, lai2021policy, binz2022heuristics, tavoni2022human}. While our computational account resonates strongly with this set of ideas, two qualifying points call for consideration. First, a great deal depends on the exact nature of the computational bottleneck hypothesized. At the center of our account is a measure related to algorithmic complexity \citep{grunwald2007minimum, hinton1993keeping, binz2022heuristics}, a measure that differs from the mutual information constraint that has provided the usual focus for value-complexity tradeoff theories \citep{lai2021policy, lerch2018policy} (see Methods). Second and still more important, the MDL-C framework does not anchor on the assumption of fixed and insuperable resource restrictions. The relevant limitations on complexity are regarded not as inherent to neural computation, but rather as advantageous for representation learning and generalization \citep{chater2003simplicity}. Indeed, while reward-complexity tradeoff models typically involve a single bottlenecked processing pathway \citep{binz2022heuristics, lai2021policy}, MDL-C includes a second pathway that  allows the agent to work around constraints on computational capacity.  This allows  for the formation of expressive, task-specific representations alongside more compressed representations that capture shared structure across tasks  \citep{musslick2021rationalizing}. 

\subsection*{Discussion}

Dual-process structure appears ubiquitously across multiple domains of human decision making. While this has long been recognized by psychological and neuroscientific models, only recently has the normative question been raised: Can dual-process control be understood as solving some fundamental computational problem? The present work has proposed an answer to this question. Starting from the problem of behavioral generalization and leveraging the concepts of compression and regularization, we derived MDL-C, a regularized version of reinforcement learning which  centers on the notion of minimum description length. Although developed from first principles, MDL-C turns out to provide a compelling normative explanation for dual-process structure. 

The account we have presented is also distinctive for its unifying character. Although sophisticated dual-process models have been proposed within each of the  behavioral domains we have considered in the present work --- executive control (e.g., \cite{lieder2018rational}), reward-based decision making (e.g., \cite{daw2005uncertainty}), and JDM (e.g., \cite{binz2022heuristics}) --- to our knowledge MDL-C is the first computational proposal to offer a unified account for empirical phenomena spanning all three of these fields. As we have noted, MDL-C also provides a novel explanation for the involvement of common neural substrates across the three relevant domains. Our treatment of the neuroscientific issues has, of necessity, been quite broad; important next steps for developing the theory would, for example, be to address regional distinctions within prefrontal cortex (see, e.g., Figure 2A-B), and  to accommodate data suggesting a multi-level hierarchy within the same region \citep{badre2018frontal}. However, even at its current granularity, MDL-C offers guidance for making a novel interpretation of a wide range of neuroscientific observations, spanning ostensibly distinct information-processing domains.

Beyond psychology and neuroscience, MDL-C bears a number of important links with existing work in machine learning and AI. In particular, it belongs to a broad class of RL systems that employ regularized policy optimization, where the agent policy is regularized toward some reference or default (see \citep{tirumala2020behavior}). Most relevant are approaches where the default policy is itself learned from experience \citep{galashov2019information, goyal2018infobot, teh2017distral, moskovitz2021towards}. In previous work involving such learning, it has been deemed necessary to stipulate an `information asymmetry,' imposing some hand-crafted difference between the observations available to the control and default policies \citep{galashov2019information, goyal2018infobot, teh2017distral, piray2021linear}. MDL-C allows this information asymmetry itself to be learned, as our simulations have demonstrated (see Figures 2C, 3B, 5E, 6D). 
Given this point and others, we are hopeful that MDL-C may prove useful to AI research, alongside psychology and neuroscience. Even in the AI context it may still be relevant that MDL-C captures aspects of prefrontal function, since PFC computations appear crucial in many settings where humans still outperform AI \citep{russin2020deep}. 

\newpage

\paragraph{Code availability.} Simulation code will be open-sourced to align with publication. 

\paragraph{Data availability.} Not applicable.

\paragraph{Acknowledgements.} We are grateful to Greg Wayne, Sam Gershman, Alex Pouget, Athena Akrami, Joe Paton, Chris Summerfield, Marcel Binz, D.J. Stouse, Dhruva Tirumala, Nathaniel Daw, and Zeb Kurth-Nelson for useful discussion. 



\bibliography{arxiv.bib}
\pagebreak


\begin{center}
  {\Large A Unified Theory of Dual-Process Control} \\ [0.8em]
    {\Large Online Methods} \\ [0.8em]
    {Ted Moskovitz \quad Kevin J. Miller \quad Maneesh Sahani \quad Matthew M. Botvinick}
\end{center}

\section{Architecture and Learning Algorithm}
All experiments employed a common architecture and learning algorithm with minor implementational variations across simulations. Our implementation of MDL-C consists of two recurrent neural networks (RNNs) which we call the \emph{control policy network} $RNN_\pi$ and the \emph{default policy network} $RNN_{\pi_0}$. These RNNs have identical sizes and architectures. They are also provided with identical inputs for each time step of experience, consisting of a one-hot encoding of the previous action, a scalar indicating the previous reward, and a vector of task-specific information (the ``observation'') which will be described separately for each task.

The control policy network $RNN_\pi$ produces as output a vector of \textit{policy} logits $\pi$ which determine the probability of taking each available action, as well as a scalar \textit{value} estimate of its expected future reward from the current state. It is trained using the \textit{advantage actor-critic} (A2C; \citep{mnih2016a2c}) algorithm, which encourages it to produces actions which maximize expected long-term reward $\mathbb E_\pi[R]$. The control policy network is also regularized using a term which encourages its action probabilities to match those produced by the default policy network. This is equivalent to encouraging the conditional description length $L(\pi|\pi_0)$ to be low. 

The default policy network also produces as output a vector of policy logits $\pi_0$. In the intact system, these are overwritten by the control policy network (see Supplementary Discussion), and so serve primarily to regularize the control policy. The default policy network is trained by \textit{policy distillation} \citep{hinton2015distilling,rusu2015policydistill} to match the output of the control policy network $\pi$. It is regularized using \textit{variational dropout} (VDO; \citep{kingma2015vdo}), which encourages its absolute description length $L(\pi_0)$ to be low.  


The overall MDL-C objective for $\pi$ and $\pi_0$ can be written 
\begin{align*}
    \mathcal L_{\mathrm{MDL-C}} = \mathbb E_\pi[R] - [\alpha D_{KL}(\pi(a|x_t;\theta)|| \pi_0(a|x_t; w)) + \beta \bar D_{KL}(q(w; \phi)|| p(w))],
\end{align*}
corresponding to the reward maximization, complexity, and goodness-of-fit terms introduced in Equation (2). Note that this expression introduces an additional weighting parameter relative to Equation (2), the rationale for which is presented in our Supplementary Discussion. Also, as will become clear in what follows, the overall objective above can be decomposed and sub-parts used to train different sectors of our agent, since only certain terms affect different pathways. Below, we describe each term in the objective in detail. 

\paragraph{Control Policy Network} Unless otherwise noted, the control policy $RNN_{\pi}$  was trained via a modification of Advantage Actor-Critic (A2C), which is described in detail in \cite{mnih2016a2c,wang2018metapfc}. Briefly, A2C is an on-policy actor-critic algorithm which weights gradients by a Monte-Carlo estimate of the advantage at each time step. In order to prevent premature convergence to suboptimal local maxima, \cite{mnih2016a2c} add an entropy bonus to the objective to prevent the policy from becoming overly deterministic early in training. In MDL-C, entropy regularization is replaced with a Kullback-Leibler (KL) divergence penalty with respect to the default policy distribution $\pi_0$:
\begin{align*}
    \nabla \mathcal L_\pi &= \nabla \mathcal L_{A2C} + \alpha \nabla\mathcal L_{KL}, \quad \mathrm{where} \\
        \mathcal L_{A2C} &= -\delta_t(x_t; \theta_v) \log \pi(a_t|x_t; \theta) + \frac{\alpha_v}{2}\delta_t(x_t; \theta_v)^2, \\
        \mathcal L_{KL} &= D_{KL}(\pi(a|x_t;\theta)|| \pi_0(a|x_t; w)), \\
    \delta_t(x_t; \theta_v) &= R_t - v(x_t; \theta_v), \\
    R_t &= \sum_{i=1}^{k-1} \gamma^i r_{t+i} + \gamma^k v(s_{t+k}; \theta_v),
\end{align*}
where $x_t = [s_t, a_{t-1}, r_{t-1}]^\top$ is the observation vector at time $t$ consisting of the state $s_t$, previous action $a_{t-1}$, and previous reward $r_{t-1}$, $\theta$ are the control policy parameters, $\alpha_v$ is a hyperparameter controlling the weight on the value-learning loss, $\theta_v$ are the value function parameters, $D_{KL}(q||p) = \sum_a q(a)\log\frac{q(a)}{p(a)}$ is the KL divergence between distributions $q$ and $p$, $w$ are the sampled parameters of the default policy network (details below), and $\gamma$ is a scalar discount factor. Intuitively, the control policy network is trained to maximize reward while simultaneously being encouraged to remain close to the behavior encoded by the default policy $\pi_0$. Early in training, obtaining reward may be challenging, and so the control policy network is primarily taught via learning signals generated by the default policy network. If the default policy network encodes behavior that is useful for the task, then learning from it may enable the control policy network to obtain reward earlier, accelerating training. In multitask settings where $\pi_0$ is conserved across tasks, it is therefore important that $\pi_0$ encodes behavior which is generally useful for the tasks faced by the agent. 

\paragraph{Default Policy Network} The default policy was trained off-policy via distillation \citep{hinton2015distilling} from the control policy network (in other words, the default policy aims to match the control policy distribution) offset by a regularization penalty on the effective bit length encoding of the network parameters:
\begin{align*}
    \nabla \mathcal L_{\pi_0} &= \nabla \mathcal L_{\mathrm{distill}} + \nabla \mathcal L_{\mathrm{VDO}} \\ 
    &= \sum_{k=1}^M \nabla_{\phi} D_{KL}(\pi(a|x_k; \theta)|| \pi_0(a| x_k; w=f(\phi; \epsilon)) + \beta \nabla_{\phi} \bar D_{KL}(q(w; \phi)|| p(w)); \\
    f(\phi^{(i)}; \epsilon) &= \phi_0^{(i)} (1 + \sqrt{\phi_\alpha^{(i)}} \epsilon^{(i)}); \quad \epsilon^{(i)} \sim \mathcal N(0, 1),
\end{align*}
where $M$ is the minibatch size of data sampled from an experience replay buffer \citep{mnih2015humanlevel}, $w$ are default policy parameters sampled using the reparameterization trick \citep{kingma2015vdo}  to allow for automatic differentiation, $\beta$ is a scalar hyperparameter weighting the regularization, and $\phi =\{\phi_0, \phi_\alpha\}$ are learned parameters defining the distribution over default policy parameters: $q(w;\phi) = \prod_{i} \mathcal N\left(w^{(i)}; \phi_0^{(i)}, \phi_\alpha^{(i)} (\phi_0^{(i)})^2\right)$, where the superscript $(i)$ denotes the $i$th parameter, and $p(w)$ is the log-uniform prior $p(|w^{(i)}|) \propto 1/|w^{(i)}|$. The noise $\epsilon$---and therefore, effectively, the default policy weights $w$---are re-sampled after every episode of training. It's this particular form of noise which limits the effective capacity of $RNN_{\pi_0}$. We use the average KL, 
\begin{align*}
    \bar D_{KL}(q(w; \phi)|| p(w)) = \frac{1}{N} \sum_{i=1}^N D_{KL}(q(w^{(i)}; \phi^{(i)}) || p(w^{(i)})).
\end{align*}
The regularization loss is computed and minimized using \emph{variational dropout} (VDO; \citep{kingma2015vdo,molchanov2017variational}), which uses a local reparameterizaztion trick to implement this KL regularization as a particular form of multiplicative noise placed on the network weights. Regularizing with respect to this choice of prior has the effect of limiting the effective bit-length of the parameters of $RNN_{\pi_0}$, reducing its effective complexity \citep{kingma2015vdo}.
Note that the distillation loss  $D_{KL}(\pi(a|x_k; \theta)|| \pi_0(a| x_k; w=f(\phi; \epsilon))$ is a direct measure of goodness-of-fit $L(\pi, \pi_0)$ (Equation 2)---the degree to which $\pi_0$ is able to match the behavior of $\pi$. To see this, note that minimizing this KL is equivalent to performing maximum likelihood estimation for $\pi_0$ with $\pi$ defining the 'true' underlying data distribution. Also observe that $RNN_\pi$ is trained on-policy and $RNN_{\pi_0}$ is trained off-policy from a buffer of experience collected by the control policy. We can view this as the control policy actively learning via trial and error in the world, while intuitively, within a multitask context, the default policy is trained to capture the behavior of the control policy on each task. The default policy thereby learns an `average' of the behaviors required to perform well on each task. However, when only a few tasks have been observed, the default policy can `overfit' to the behaviors learned on those initial tasks. This can be problematic, as if future tasks differ, the default policy could misguide the learning process for the control policy (see above). The VDO complexity regularization forces the default policy network to simplify, preventing overfitting and facilitating generalization.

\paragraph{Architecture Details} The LSTM dynamics were governed by the following standard equations:
\begin{align*}
    i_t &= \sigma(W_{xi} x_t + W_{hi} h_{t-1} + b_i) \\
    f_t &= \sigma(W_{xf} x_t + W_{hf} h_{t-1} + b_i) \\
    o_t &= \sigma(W_{xo} x_t + W_{ho} h_{t-1} + b_o) \\
    c_t &= f_t \odot c_{t-1} + i_t \odot \tanh\left(W_{xc}x_t + W_{hc} h_{t-1} + b_c \right) \\
    h_t &= o_t \odot \tanh(c_t),
\end{align*}
where $i_t$, $f_t$, $o_t$, $c_t$, $h_t$ are the input gate, forget gate, output gate, cell state, and hidden state at time $t$, respectively, $\sigma(x)=1/(1 + \exp(-x))$ is the sigmoid function, and $\odot$ denotes element-wise multiplication. In order to assess the degree to which the default policy learned to ignore certain input features, an element-wise gating layer $\ell(x)$ was applied to the input:
\begin{align*}
    \ell(x_t) = \sigma(\tau \omega) \odot x_t,
\end{align*}
where $\tau$ was a hyperparameter fixed across simulations and $\omega$ was a learned vector of parameters with dimension equal to that of the input. As $\omega_d\to\infty$, the $d$th input feature is passed on to the layers above, while if $\omega_d\to-\infty$, the $d$th input feature is gated out. Importantly, gradients from the VDO loss (see above) did not flow into $\omega$, only those from the distillation loss, so $\omega$ learned to gate features in or out that were already either being used or ignored by the network, rather than simply being ablated directly as the VDO penalty increased. We found that adding this gating layer did not affect the performance of the agent. 


\paragraph{Training Details}
In all simulations, the agent was updated using a learning rate of $\eta=0.0007$, a value function loss weight of $\alpha_v=0.05$, a policy KL weight of $\alpha=0.1$, a discount factor of $\gamma=0.9$, a gating layer coefficient of $\tau=150$, and a VDO KL weight of $\beta=1.0$ unless otherwise noted. All gradient updates were performed using the Adam optimizer \citep{kingma2014adam}. To generate the plots in Figure 4, agents were trained on their respective simulations with VDO KL weight $\beta\in \{0.0, 0.1, 1.0, 10.0, 100.0\}$ for all tasks except the two-step task, for which $\beta\in\{1.0, 10.0, 100.0, 1000.0, 10,000.0\}$ was used. After training, the average KL between policies and average VDO complexity KL were computed over 100 trials with frozen weights. Total KL was computed as the sum of these two quantities, with the VDO KL scaled by $\beta=100$ for the two-step task. Further simulation-specific details can be found below, and a summary of hyperparameter values for each task is provided in Table \ref{table:hyperparams}. All results were obtained by averaging over 8 random seeds, with shading on line plots denoting one unit of standard error. 

\begin{table}[ht]
  \begin{center}
  \begin{tabular}{lr}
  \toprule
   \textbf{Task}    &   ($\eta$, $\alpha_v$, $\alpha$, $\beta$ $\gamma$, $M$, $\tau$, \# hidden, $\pi_0$ buffer size)  \\
  \midrule
   Navigation        &    (7e-4, 0.05, 0.1, 1.0, 0.9, 32, 150, 48, 1e5)  \\
   Stroop           &  (7e-4, 0.05, 0.1, 1.0, 0.9, 1, 150, 48, 1e5)  \\
   Demand Stroop           &   (7e-4, 0.05, 0.1, 1.0, 0.9, 1, 150, 48, 1e5) \\
   Two-Step          &     (7e-4, 0.05, 0.1, 3.0/100.0$^*$, 0.9, N/A, 150, 48, N/A) \\
   O \& D &   (7e-4, 0.05, 0.1, 1.0, N/A, 1, 150, 5, 1e3) \\
   Continuous Control      &  (3e-4, 0.05, 0.1, 1.0, 0.99, 128, 150, 256, 1e6) \\
   Heuristics    & (3e-4, N/A, 0.1, 1.0, N/A, 32, 150, 128, 1e6) \\
  \hline
  \end{tabular}
  \end{center}
  \caption{Hyperparameters for each task. $^*$The Two-Step task settings are described in greater detail below.}
  \label{table:hyperparams}
  \end{table}

\section{Generalization}
\paragraph{Navigation} In this experiment, the agent was trained on two tasks within the classic FourRooms environment \citep{sutton1999options}, an 11$\times$11 gridworld in which the available actions were\\ $\{\texttt{up}, \texttt{right}, \texttt{down}, \texttt{left}\}$. The agent LSTMs in both tasks each had 48 hidden units. Input feature importance to $RNN_{\pi_0}$ was tracked using an element-wise gating layer (see ``Architecture Details''). The state input at each time step $s_t$ was an 11-dimensional vector comprising the index of the agent's current state, a flattened $3\times 3$ square representing the agent's immediate surroundings (specifically, a value of 0 indicated a free space and a value of -1 indicated a barrier), and the index of the current goal location. Agents were trained in the following way: at the start of each episode, a goal location was randomly sampled from a set of possible goal locations. Episodes terminated when the agent either reached the goal location or once 250 steps had elapsed. The agent updated its weights after each episode. In the first task (Figures 1B, 2C), training was broken into two phases. In the first phase, the goal for each episode was randomly sampled from a set of possible goal states which were arranged in a checkerboard pattern (see Figure 1B inset). The agent was trained in this setting for 40,000 episodes. In the second phase, the checkerboard pattern was inverted, so that candidate goal states from the first phase could no longer contain a goal and vice versa. The agent was trained in this phase until it was able to reach a second phase goal in the fewest possible number of steps. The MDL-C agent was compared against a standard regularized policy optimization (Standard RL) agent with the same design, except it did not have the complexity KL penalty on its $RNN_{\pi_0}$. 

In the second task (Figure 2D), at the start of each episode, a goal location was randomly sampled as either the top left state with probability 90\% or the bottom right state with probability 10\%, with the agent starting anywhere in the environment with uniform probability. The agent was trained for 30,000 episodes. 

\paragraph{Continuous Control} In this setting, the agent was trained sequentially on tasks from the Walker domain from the DeepMind control suite \citep{tass2018dmc}.  In this multitask setting, tasks were sampled one at a time uniformly without replacement from the available tasks within the Walker domain, with $\pi_0$ conserved across tasks. For Walker (visualized in Figure 1C inset), these tasks are \texttt{stand},  \texttt{walk}, and \texttt{run}. Observations are 25-dimensional feature vectors, with 1 dimension encoding the height of the agent's center of mass, 14 dimensions encoding position, 1 dimension encoding reward, and 9 dimensions encoding the velocity of various components of the agent. The action space is a 6-dimensional continuous vector which directs the agent's joints. In \texttt{stand}, reward is granted in proportion to the agent's height, and in \texttt{walk} and \texttt{run} further reward is given for forward velocity. See \cite{tass2018dmc} for further environment details. Performance results for the \texttt{run}, the hardest task in the Walker domain are plotted in Figure 1C, where $k$ indicates the task round at which the task was sampled. Unlike in other simulations, here $\pi$ was trained off-policy using \emph{soft actor-critic} (SAC; \citep{haarnoja2018sac}) in order to improve sample-efficiency. As baselines, we included standard SAC (see \cite{haarnoja2018sac} for details) and a MDL-C agent without the VDO complexity KL cost (termed RPO). Agents were trained on each task for 250,000 environment steps. As $k$ increases---in other words, as the agent has been trained on more tasks---MDL-C's performance improves substantially. 

\section{Simulation 1: Executive control}

\paragraph{Stroop Task} In the Stroop task, the agent must perform either ``word-reading'' (WR) or ``color-naming'' (CN) tasks across two different colors and two different words, totalling eight different possible stimuli: (\textcolor{red}{red}[WR], \textcolor{blue}{blue}[WR], \textcolor{red}{blue}[WR], \textcolor{blue}{red}[WR], \textcolor{red}{red}[CN], \textcolor{blue}{blue}[CN], \textcolor{red}{blue}[CN], \textcolor{blue}{red}[CN]), each presented to the agent as a three-dimensional vector $x_t=[\texttt{color},$ \newline $\texttt{word}, \texttt{task}]$ with the following encodings: blue$\to-1$, red$\to+1$, CN$\to-1$, WR$\to+1$. The presentation frequencies were 20\% for all WR stimuli and 5\% for all CN stimuli. There were two possible actions, corresponding to $-1$ and $+1$. The agent received $+1$ reward when its action matched the appropriate value of the stimulus feature (e.g., if the task feature is $-1$ and the color feature is $+1$, the desired action is $+1$) and zero otherwise. In order to simulate reaction times (RTs), the input stimulus for a given trial was re-presented up to a maximum of 5 times to the agent until the entropy of the control policy $H[\pi] = -\sum_{a}\pi(a|x_t;\theta)\log\pi(a|x_t;\theta)$ dropped below a threshold  $b=0.5$, similar to the approach to modeling RTs used by \cite{cohen1990stroop,song2016rt}. The RT for each trial was the number of presentations of the stimulus until a response was produced. After a response was generated, the trial ended. The agent was trained for 15,000 trials and each LSTM had 48 hidden units. 

\paragraph{Demand Avoidance Stroop Task} 
In this task, the agent was presented with word-reading (WR) and color-naming (CN) trials, encoded as in the Stroop task described above, with each WR stimulus having a 20\% chance of presentation on any given trial. CN trials were presented 20\% of the time, but in this task CN trials consisted of two time steps. On the first time step, the agent was presented with the stimulus $[0, 0, -1]$ to indicate a CN trial. Its action at this stage served to select between two categories, referred to as \emph{high-demand} and \emph{low-demand}. If the agent selected the high-demand category, then in the second time step of the trial, a conflict stimulus was presented with a 90\% chance and a congruent stimulus with a 10\% chance. If the agent selected the low-demand category, these probabilities were reversed. The agent shared the same settings as in the main Stroop task, and was also trained for 15,000 trials. 

\paragraph{Interference Stroop Task}
In this case, the agent was first pre-trained on color-naming-only trials for 30,000 trials. It was then trained for 45,000 trials on word-reading. During this training phase, the agent's reaction time was evaluated on both CN and WR trials (in evaluation trials, the agent's weights were not updated). The agent architecture and environment set-up were the same as in the ``standard'' Stroop task above. One important characteristic of the Stroop task is that there are interaction effects between task and congruity---that is, at the outset of training, color interferes with shape more than shape interferes with color. This means that shape-naming conflict trials result in disproportionately higher RTs compared to shape-naming congruent trials compared to the difference between color-naming conflict trials and color-naming congruent trials. This relationship is then reversed at the end of training, with a greater relative increase for color-naming trials. We verified that this property held for MDL-C by running a one-way ANOVA test on the differences in RT between each trial type (e.g., color-naming conflict RT - color-naming congruent RT vs. shape-naming conflict RT - shape-naming congruent RT), with the interaction effect at 0 trials yielding $F = 11.3$ and $p = 0.005$ and the interaction effect at 44,000 trials yielding $F = 13.7$ and $p = 0.002$.

\paragraph{Zero-Shot Stroop Task}
The agent was trained 8,000 trials in which the unneeded feature was zeroed out from the stimulus (i.e., the agent gets 3d inputs [color, word, task id], where -1 = blue, +1 = red in the first two dims, -1 = color-naming, +1 = word-naming for task id, and 0 = NULL in any location). So, the agent would see [-1, 0, -1] for a "blue" color-naming task. The stimulus distribution was uniform. During training, both MDL-C and `Regular RL'' (just a control policy $RNN_\pi$ with otherwise identical hyperparameters) get to 100\% accuracy. The agent is then evaluated on 100 trials with fixed weights in which the unneeded feature is included in the stimuli. The evaluation performance is the percent correct over 100 evaluation trials with fixed weights. To test the hypothesis that the improved performance of MDL-C is rooted in robustness to changes in inputs, we also measured the average KL between the policy distributions for each approach on masked inputs (like the ones on which they were trained) and on inputs with the missing feature included. Regular RL had a greater difference, indicating that responses were more effected by the out-of-distribution inputs.

\section{Simulation 2: Reward-based learning}

\paragraph{Two-Step Task}
We use a variant of the two-step task based on the one used by \cite{wang2018metapfc} in which transition contingencies--in addition to reward contingencies--may switch. The task was changed in this way following the finding by \cite{akam2015twostep} that when transition contingencies are fixed, a habit-like strategy in which second stage states which have recently yielded reward are directly mapped to actions in the choice stage can develop which closely matches the pattern of behavior expected of agents using planning. 
Additionally, the agent is provided with an input feature which indicates which transition contingency setting is currently active (an ingredient added to the task from \cite{akam2015twostep} in order to restore the property that model-based and -free strategies yield the classical patterns shown in Figure 5C) . To use this feature to inform its actions, the agent must compute a higher-complexity policy than if this feature is ignored, analogous to the difference between classifying inputs according to \texttt{XOR} versus \texttt{OR} logic. 
To be more precise, with two second stage states $A$ and $B$ and two actions $a_L$ and $a_R$, we can have either
\begin{align*}
\text{Setting 0} &= \begin{cases}
p(A|a_L) = 0.8, & p(B|a_L) = 0.2 \\
p(A|a_R) = 0.2, & p(B|a_R) = 0.8
\end{cases}
\end{align*}
\begin{align*}
\text{Setting 1} = \begin{cases}
p(A|a_L) = 0.2, & p(B|a_L) = 0.8 \\
p(A|a_R) = 0.8, & p(B|a_R) = 0.2
\end{cases}
\end{align*}
In other words, in one setting $a_L$ is likely to lead to $A$ and $a_R$ is likely to lead to $B$, and in the other, the reverse is true. The agent is shown a binary feature which indicates which transition setting the environment is in (however, it has to learn what this feature means through experience). More precisely, the state observation at each time step $s_t$ is a 5-dimensional vector, with the first four dimensions comprising a one-hot encoding of the current position of the agent within the task (either \texttt{fixation stage}, \texttt{choice stage}, \texttt{A}, or \texttt{B}), with the final dimension a binary encoding of the current transition setting. The agent has three possible actions: $a_L$, $a_R$, and $a_{fixate}$, which the agent is required to produce in order to progress from the fixation stage to the choice stage. There are also two possible settings for the reward contingencies, with either $A$ or $B$ having a 90\% chance of leading to reward, with the other state in either contingency having a 10\% chance. The agent is trained for 16,000 episodes, where each episode consists of 100 trials. At the end of each episode, the agent networks' hidden states are reset and an update is performed via backprop. On any given trial, there is a 2.5\% chance that the reward contingency switches and a 5\% chance that the transition contingency changes. During training, we found it helpful to start with a 0\% chance of reward contingency switches and linearly increase the probability to 2.5\% over the first 2,000 episodes, as this helped the agent reliably learn the meaning of the transition setting feature. All other task settings and analysis details for stay probabilities and logistic regression are the same as in \cite{wang2018metapfc}. Importantly, in this task the default policy was trained online (but still off-policy) via full trajectories collected by the control policy, rather than via a buffer of $(s,a,r,s')$ tuples. This is because the full episode history is required to effectively meta-learn, as demonstrated by \cite{wang2018metapfc,wang2017l2rl}. The hyperparameter settings used to generate the plots in Figure 5(D-F) were identified after an initial grid search with eight random seeds per $(\alpha, \beta, RewardScale)$ tuple with $\alpha \in \{0.05, 0.1, 0.2\}$, $\beta\in\{0.1, 1.0, 3.0, 5.0, 10.0, 100.0\}$, $RewardScale \in \{0.5, 0.75, 1.0\}$ and further confirmed by an additional eight random seeds, for a total of 16. The `classic' MB-MF patterns were obtained with $(0.1, 100.0, 1.0)$ and the mixed patterns were observed with $(0.2, 3.0, 0.75)$. To further support the mixed MB-MF-ness of the response pattern in Figure 5F, we performed Wilcoxon signed-rank tests between the average of the rewarded, common and unrewarded, uncommon responses and the average of the rewarded, uncommon and unrewarded, common responses as a measure of model based-ness, and between rewarded, common and rewarded, uncommon responses and unrewarded, common and unrewarded, uncommon responses as a measure of model free-ness. The response patterns for both the control and default policies were statistically significant ($p = 0.012$) for both model-based and model-free behavior. The agent's LSTMs had 48 hidden units each.

\paragraph{Perseveration}
In this experiment, the agent was trained on the drifting two-armed bandit task from \cite{miller2018predictive}. In this task, trials consist of a single time-step in which the agent has two possible actions, with the probability of reward for each arm evolving with a Gaussian random walk. Specifically, if the probability of being rewarded by choosing a given action on trial $t$ is $P_t$, then the probability of being rewarded for choosing that arm on the next trial is sampled from the distribution $P_{t+1} \sim \mathcal N(P_t, 0.15^2)$. The agent either receives a reward of 1 or 0, and is trained for 3,000 trials. In this case, each RNN had 5 hidden units. After training, logistic regression is performed to predict the agent's behavior on a given trial, with the regressors being the choice made at each time-step ($\pm1$, the whether a reward was given at each time-step ($\pm1$), and their product. A high regression weight for previous choices indicates a tendency to perseverate, a high weight for the outcome/reward indicates that the agent is influenced by whether it was rewarded at each step independent of its previous choices, and a high regression weight for their product indicates that the agent is influenced by choices that led to rewards (reward-seeking behavior). 

\paragraph{Omission and Contingency Degradation}
We use the same task set-up as \cite{miller2019habits}. As in \cite{miller2019habits}, in order to model the effect of overtraining on the agent's sensitivity to omission of reward, the agent was first trained on a two-armed bandit task in which action 1 (``lever press'') led to a reward of 1 with 50\% probability and action 2 (``leisure'') resulted in a reward of 0.1 100\% of the time. It was then trained for 750 trials on a modification of the task in which reward was never delivered for lever pressing and in which leisure resulted in a reward of 0.1 half the time and a reward of 1.1 half the time. The agent's lever-pressing probability $P(\mathrm{lever\ press})$ was then measured at the end of the second training phase. This probability was plotted against the number of trials $T$ for which the agent was trained on the first phase, where $T\in[100, 200, 300, \dots, 2000]$. The contingency degradation variant of this task was exactly the same, except that the leisure action always resulted in a reward of 0.1 in the second phase.

\section{Simulation 3: Judgment and decision-making}

\paragraph{Heuristics} 
We use the same experimental setting as \cite{binz2022heuristics}. Briefly, the agent is meta-trained on a series of randomly generated paired comparison tasks with continuous input features $x$ in which it must predict which of two presented feature vectors $x_t = (x_t^A, x_t^B)$ is associated with a higher value of an unobserved scalar criterion $y_t=(y_t^A, y_t^B)$. More precisely, for each task $i$, there is an underlying linear relationship between features and the unobserved criterion: 
\begin{align*}
    y_{t,A} &= w_i^\top x_{t,A} + \epsilon_{t,A}; \\
    y_{t,B} &= w_i^\top x_{t,B} + \epsilon_{t,B},
\end{align*}
where $\epsilon_{t,A},\epsilon_{t,B}\sim \mathcal N(0, \sigma^2)$, with $\sigma^2$ a fixed variance and $w_i \in \mathbb R^4$. An ideal observer model then expresses the probability that $y_A > y_B$ as
\begin{align}
    p(y_A > y_B|x, w_i) = p(C=1|x, w_i) = \Phi\left(\frac{w_i^\top x}{\sqrt{2}\sigma} \right), 
\end{align}
where $\Phi(\cdot)$ is the cumulative distribution function of a standard Gaussian distribution and $C\in\{0, 1\}$ is a binary random variable which evaluates to 1 when $y_A > y_B$ and 0 otherwise. Task feature weights $w_i$ are randomly generated from a standard normal distribution, and the agent is meta-trained to estimate a posterior distribution over $w$ with minibatches of 32 tasks and each task being presented to the agent for 10 trials. The reward for a given trial can be modeled as the log likelihood: $p(C_t|x_t, \phi_t, $ For a more detailed description of the training process, see \cite{binz2022heuristics}.
The control network $RNN_\pi$ produces the parameters (mean $\mu_t$ and variance $\Psi_t$) of an approximate Gaussian posterior over $w_i$, which is then integrated into a predictive distribution for classification:
\begin{align*}
    p(C_{t+1}|x_{t+1},\phi_t, \Theta) = \int p(C_{t+1}|x_{t+1}, w) q(w;\phi_t) \ dw
\end{align*}
where and $\phi_t=\{\mu_t, \Psi_t\}$ are the parameters of the approximate Gaussian posterior $q$ over parameters $w$. The conditional distribution is as above
\begin{align*}
    p(C_{t+1}=1|x_{t+1}, w) = \Phi\left(\frac{w^\top x_t}{\sqrt{2}\sigma} \right).
\end{align*}
  The default network $RNN_{\pi_0}$ also produces parameters $\phi_t^0$ of a Gaussian $q_0$ and is trained to minimize $D_{KL}(q(w, \phi_t) || q_0(w, \phi_t^0))$ in addition to the VDO complexity KL weighted by $\beta$ (see ``Default Policy'' details above). 

To test the emergence of heuristics, we use the task variant from \cite{binz2022heuristics} in which there is a known ranking of input features, which classically induces a form of one-reason decision-making termed ``take the best'' (TTB), wherein subjects make decisions based on the top-ranked feature which differs between two inputs. To measure the emergence of such a heuristic in artificial agents, \cite{binz2022heuristics} use the \emph{Gini coefficient} $G$ \citep{atkinson1970gini} measured over the feature weights $w$, defined below: 
\begin{align*}
    G(w) = \frac{\sum_{i=1}^d\sum_{j=1}^d |w_i - w_j|}{2d \sum_{i=1}^d w_i}.
\end{align*}
The Gini coefficient can be thought of as a measure of inequality among feature weightings, so that it tends to $1$ when one feature grows in importance compared to the others, and tends to $0$ as all feature weights $w_i$ converge to the same value. 
As a means of probing the effect of reducing the relative cost of employing a compensatory strategy (Figure 6D), we reduced the weighting on the KL between the default and control policies, setting $\alpha=0.01$. This effectively lowers the penalty for deviation in behavior from the capacity-limited policy.


\section{Supplementary Discussion}

\subsection*{Computational Framework}

We discuss here several details of the MDL-C computational framework.

In importing MDL into reinforcement learning, we made a link between the `data' to be compressed and the agent's policy. Several additional comments are warranted concerning this choice.  First, it is important to note that MDL, or compression more generally, can also be applied quite naturally to other data structures within RL. One obvious target for compression is the action-outcome model that lies at the center of model-based reinforcement learning, and this is just one of several candidates (see \cite{botvinick2015reinforcement}). While our MDL-C proposal focuses on the agent policy, this in no way excludes  other compression targets. 

Directing MDL toward the agent policy results in a setup that differs in some subtle and interesting ways from what is involved in classical MDL. In the latter setting, the data are typically assumed to be fixed. In MDL-C, in contrast, the data (since they comprise the agent's policy) are subject to continual change. Furthermore, while in classical MDL the data are assumed to be independent of MDL itself, the data in MDL-C can be altered over time in response to pressures that arise from the MDL objective. This feature could be avoided if MDL-C were implemented as a strict constrained optimization process, with no trade-off between reward maximization and compression. However, as it turns out, some of the empirical phenomena addressed in the main paper arise specifically from the trade-off that our implementation involves. One obvious example of this is the demand-avoidance effect described in Experiment 1, where the policy is clearly influenced by the compression terms in the MDL-C objective. An interesting target for next-step research would be to consider the possible psychological and neuroscientific implications of this hypothesized trade-off, and in particular to consider whether a clear normative justification for this feature of MDL-C might be identified. 

As discussed under Methods, the trade-off between value and compression in MDL-C is controlled by an adjustable hyperparameter. As also noted there, our neural network implementation also includes hyperparameters weighting the complexity and deviation terms against one another. This may seem surprising to readers familiar with classical MDL, where no such relative weighting occurs. However, it should be noted that the complexity term in classical MDL directly quantifies algorithmic or Kolmogorov complexity (see \cite{grunwald2004tutorial}), whereas our implementation quantifies complexity in terms of the weight distribution of a neural network. This weight distribution serves as a proxy for algorithmic complexity, since it affects the complexity of the policies the network implements. However, it is not identical, nor is it guaranteed to quantify complexity on a similar scale, thus requiring the introduction of a scaling parameter. A similar point pertains to the other KL cost in the objective function used in our implementation, capturing the divergence between policies $pi$ and $pi_0$. This, too, is a proxy for the corresponding term in classical MDL, which again is intended to capture algorithmic complexity. 

An additional aspect of our MDL-C implementation that bears further discussion is the process by which $RN\!N_\pi$ interfaces with ${RN\!N}_{\pi_0}$ at decision or inference time. In our neural network implementation, at least as the code is written, the interaction is quite straightforward: ${RN\!N}_{\pi_0}$ outputs its policy and then this is simply overwritten by $RN\!N_\pi$. This way of describing the interaction may appear to stand in tension with our description in the main text of ${RN\!N}_{\pi}$ ``overriding'' or ``endorsing'' ${RN\!N}_{\pi_0}$. However it should be noted that there is a notational variant of our implementation that aligns much better with these descriptions. Specifically, one can view ${RN\!N}_{\pi}$ as \textit{adding} or \textit{subtracting} from the action probabilities specified by ${RN\!N}_{\pi_0}$ (or, alternatively, adjusting them in a multiplicative fashion), with the result corresponding to $\pi$. If ${RN\!N}_{\pi}$ is viewed as outputting a vector of differences or deltas, then an output of zero can be interpreted as an ``endorsement'' of ${RN\!N}_{\pi_0}$, and any other output can be interpreted as ``overriding'' ${RN\!N}_{\pi_0}$.  

One final comment on the computational framework relates to the claims in the main text about generalization performance.  These may appear to stand in tension with some of the phenomena simulated in our experiments. For example, the behavioral inflexibility seen in contingency degradation after extended pre-training may appear to contradict the idea that compression, in the style of MDL-C, fosters rapid adaptation to new task challenges. However, it should be noted that the idea of `generalization' can cut both ways. Adversarial environments can be constructed where an agent's tendency to base action selection on past outcomes yields what looks like maladaptive behaviour. Contingency degradation with extensive pretraining can be seen as adversarial in this sense. The claim that dual-process organization supports generalization on average is thus reconcilable with cases where it can be understood to cause locally suboptimal behavior.

\subsection*{Results}

In the main text, we motivate MDL-C by focusing on the problem of generalization. It may seem surprising, then, that none of the dual-process phenomena addressed in our Results section involve behavioral generalization. This is, of course, not a flaw in our account. Our simulations show that dual-process phenomena can be understood as reflecting the operation of a mechanism that \textit{elsewhere and more generally} supports behavioral adaptation. This point is illustrated  by juxtaposing the results presented in Figure 2 in the main text from those presented in Figure 1B (left), since both relate to navigation.  As a stimulus to further research, it is worth describing an informal, exploratory simulation in which we investigated the role that generalization might be understood to play in one other task we addressed in our simulations, namely the Stroop task. The Stroop task can be understood as reflecting a simple form of flexible generalization: People performing the task are able, based on a verbal instruction, to ignore word identity and name colors, despite never (or at least rarely) having encountered colored \textit{color words} in a color-naming task context before. To study this kind of flexibility in MDL-C, we trained our network agent to perform color-naming on inputs indicating color but not word identity, and also to perform word-reading on inputs indicating color-word identity but not color \textit{per se}. As soon as training had proceeded far enough to yield error-free task performance, we introduced Stroop inputs including both color and color-word information. Given a task cue, the agent responded much more accurately to such inputs than a baseline agent trained without description-length regularization,  that is, using only the RL term in the MDL-C objective (data not shown). This informal result suggests that MDL-C learned to `attend' only to task-relevant input channels during the initial training, preparing it to attend selectively when faced with Stroop stimuli. It is our hope and expectation that further simulation work along lines such as these may generate further testable predictions from MDL-C in the task settings addressed in our simulations of dual-task phenomena.


\newpage
\begin{figure*}[ht]
	\begin{center}
        \includegraphics[width=0.99\linewidth]{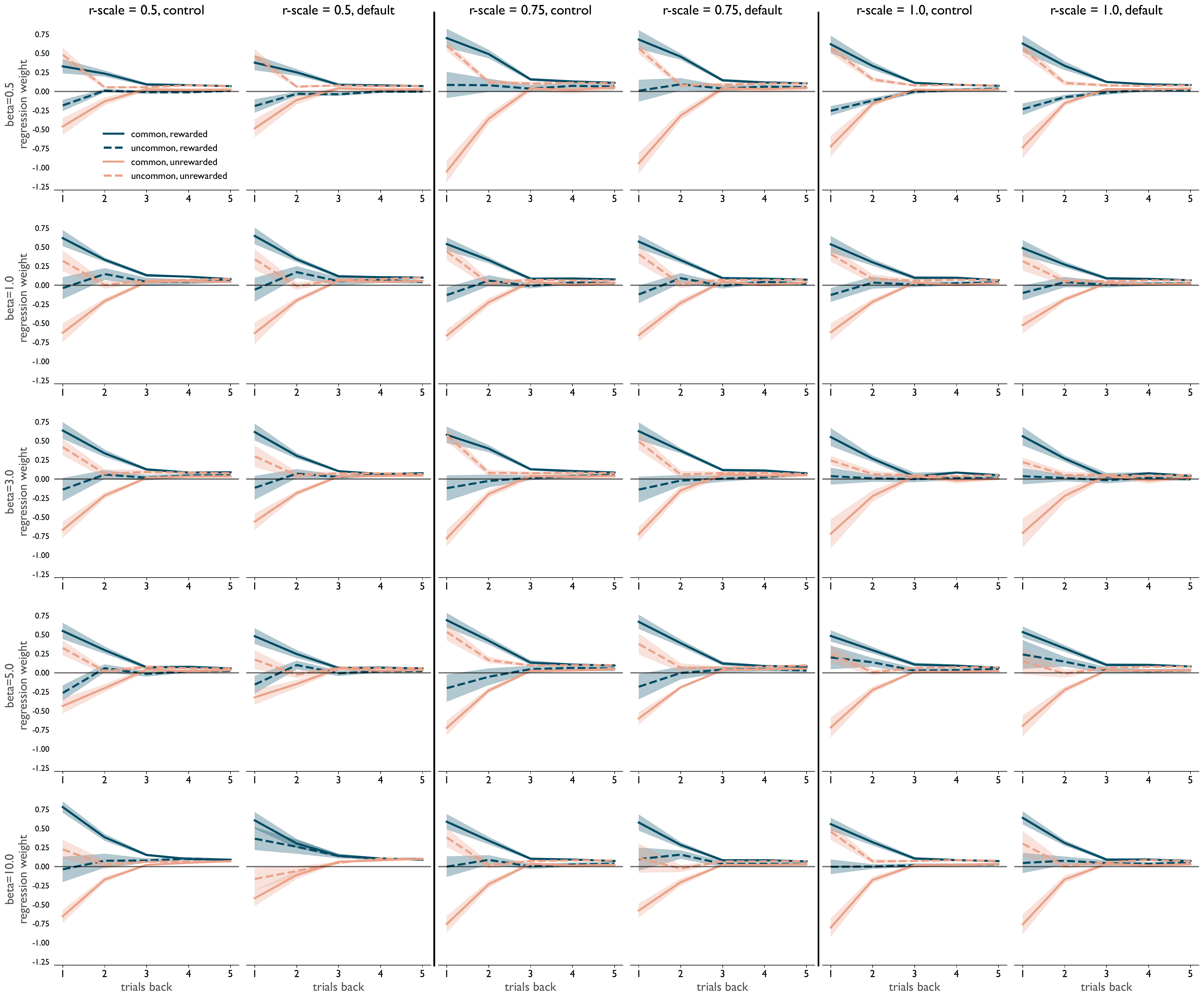}
	\end{center}
	\vspace{-2em}
	\caption{Two-step results from full hyperparameter sweep described in Methods, with $\alpha = 0.05$. Format as in Figure 5 in the main text. 
	}
	\label{fig:fig8}
	\vspace{-1em}
\end{figure*}

\begin{figure*}[ht]
	\begin{center}
		\includegraphics[width=0.9\linewidth]{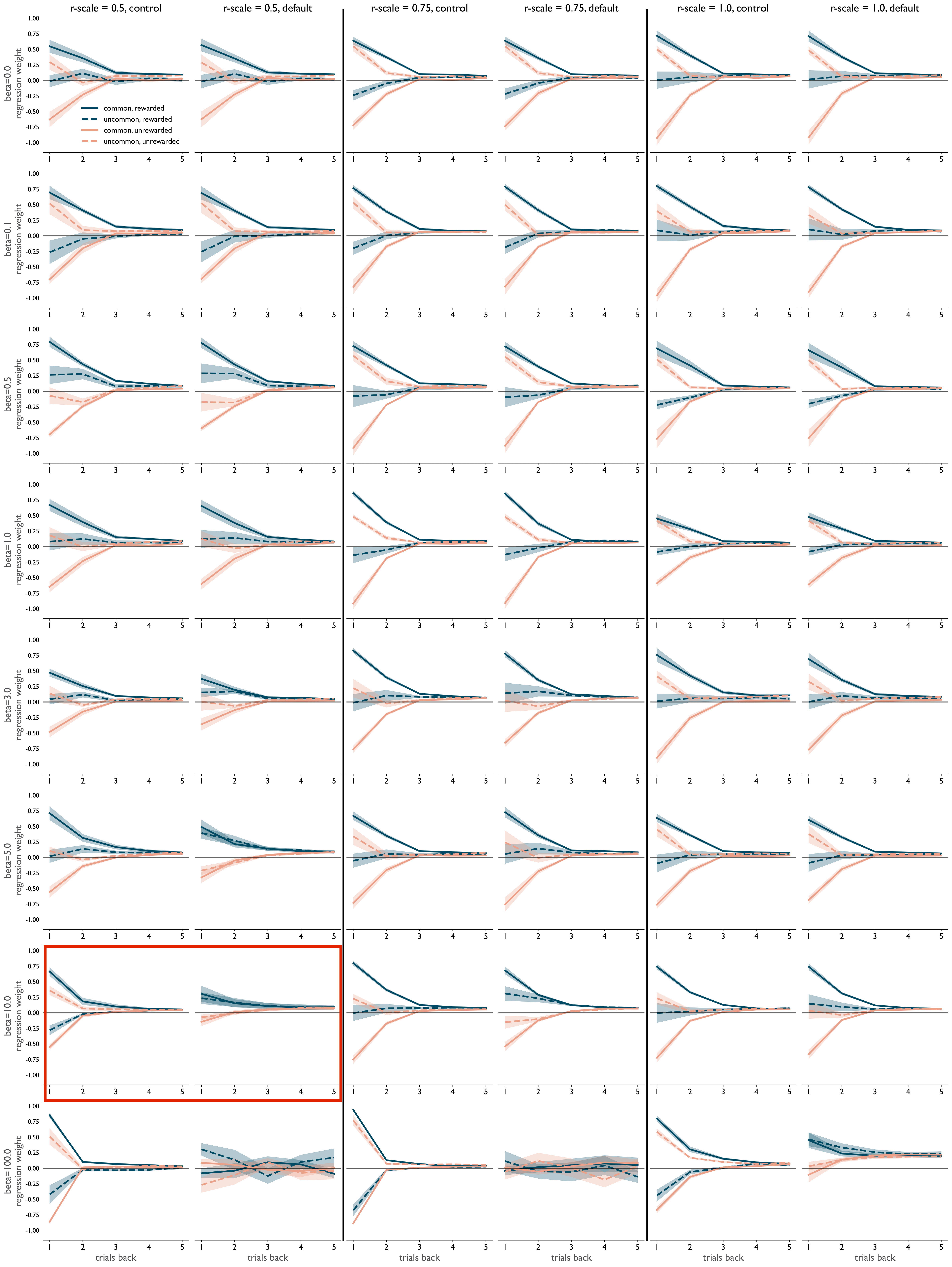}
	\end{center}
	\vspace{-2em}
	\caption{Two-step results from full hyperparameter sweep described in Methods, with $\alpha = 0.1$. The boxed plot appears in Figure 5D in the main text. Format as in Figure 5 in the main text. 
	}
	\label{fig:fig9}
	\vspace{-1em}
\end{figure*}

\begin{figure*}[ht]
	\begin{center}
        \includegraphics[width=0.99\linewidth]{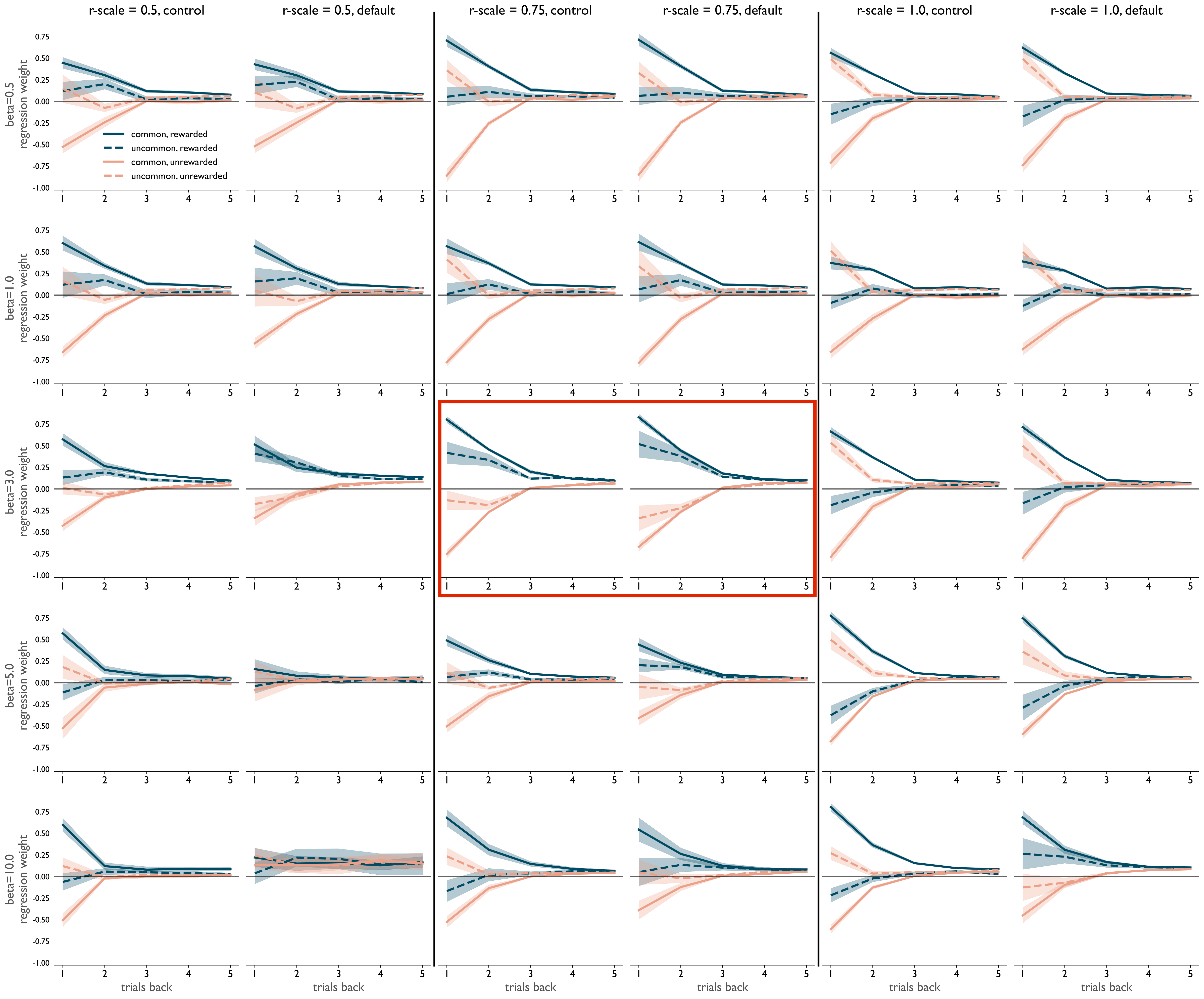}
	\end{center}
	\vspace{-2em}
	\caption{Two-step results from full hyperparameter sweep described in Methods, with $\alpha = 0.2$. The boxed plot appears in Figure 5E in the main text. Format as in Figure 5 in the main text.  
	}
	\label{fig:fig10}
	\vspace{-1em}
\end{figure*}

\end{document}